\documentclass[aip,pop,amsmath,amssymb,reprint,superscriptaddress]{revtex4-1}

\usepackage{graphicx} 
\usepackage{dcolumn}
\usepackage{bm}
\usepackage{amssymb}
\usepackage{amsmath}
\usepackage{wasysym}
\usepackage{color}
\usepackage{soul}
\usepackage{hyperref}
\usepackage[toc,page]{appendix}
\usepackage{float}
\usepackage{flushend}
\usepackage{hyperref}
\hypersetup{
    colorlinks,%
    citecolor=blue,%
    filecolor=blue,%
    linkcolor=blue,%
    urlcolor=blue
} 

\begin{document}

\preprint{AIP/123-QED}

\title{Dynamo transition in a five-mode helical model}

\author{Rohit Kumar}
\thanks{Present address: Institut de Recherche en Astrophysique et Plan{\'e}tologie, Universit\'e de Toulouse, 14 avenue Edouard Belin, 31400 Toulouse, France}
\email[Email:]{rohitkumar.iitk@gmail.com}
\affiliation{Department of Physics, Indian Institute of Technology, Kanpur 208016, India}
\author{Pankaj Wahi}
\email[Email:]{wahi@iitk.ac.in}
 \affiliation{Department of Mechanical Engineering, Indian Institute of Technology, Kanpur 208016, India}

\begin{abstract}
We construct a five-mode helical dynamo model  containing three velocity and two magnetic modes and solve it analytically.  This model exhibits dynamo transition via supercritical pitchfork bifurcation.   We show that the critical magnetic Reynolds number for dynamo transition ($\mathrm{Rm}_c$) asymptotes to constant values for very low and very high magnetic Prandtl numbers ($\mathrm{Pm}$). Beyond dynamo transition,  secondary bifurcations lead to periodic, quasi-periodic, and chaotic dynamo states as the forcing amplitude is increased and chaos appears through a quasi-periodic route.
\begin{description}
\item[PACS numbers]{47.35.Tv, 47.65.-d, 05.45.Ac}
\end{description}
\end{abstract}

\maketitle

\section{introduction}

The generation of magnetic field in stars, planets, and galaxies is explained by dynamo mechanism. \cite{Moffat:book} Dynamo transition has been studied in experiments, direct numerical simulations, and through theoretical analysis.~\cite{Monchaux:PRL2007,Ponty:PRL2004,Kazantsev:JETP1968} The nature of dynamo depends on magnetic Prandtl number $\mathrm{Pm} = \nu/\eta$, \cite{Brandenburg:PR2005} where $\nu$ is the kinematic viscosity and $\eta$ is the magnetic diffusivity of the fluid. The magnetic Prandtl number for stellar and planetary dynamos is very small, whereas for galactic systems, it is very large.~\cite{Plunian:PR2013} In this paper, we present a low-dimensional model to understand the dynamo transition for limiting cases of very high and very low magnetic Prandtl numbers.

Various aspects of dynamos have been studied using direct numerical simulations (DNS), \cite{Glatzmaier:NATURE1995,Ponty:PRL2007,Mininni:PRE2005,Ravelet:PF2005,Yousef:PRL2008,Yadav:EPL2010,Kumar:EPL2013,Kumar:JT2015} which is one of the most powerful tools to understand complex natural phenomena especially when it is difficult and expensive to get meaningful experimental data as in the case of the dynamo effect. In numerical simulations two kinds of Reynolds numbers are defined: kinetic Reynolds number $\mathrm{Re} = UL/\nu$ and magnetic Reynolds number $\mathrm{Rm} = UL/\eta$, where $L$ and $U$ are respectively the large-scale length and velocity of the system. These dimensionless parameters are related as $\mathrm{Rm} =\mathrm{Re} \, \mathrm{Pm}$. In numerical simulations an external force field is applied to a conducting fluid in the presence of a small seed magnetic field and dynamo transition is observed. Results indicate that the dynamo transition occurs above a critical magnetic Reynolds number, $\mathrm{Rm}_c > 1$.~\cite{Ponty:PRL2005,Haugen:PRE2004} 

Several research groups have investigated dynamo transition using DNS and reported that the dynamo exists for both the low and the high magnetic Prandtl numbers.~\cite{Schekochihin:APJ2004,Schekochihin:PRL2004b} Ponty {\it et al.}~\cite{Ponty:PRL2005} studied the dependence of $\mathrm{Rm_c}$ on $\mathrm{Pm}$ by performing low-$\mathrm{Pm}$ dynamo simulations. They observed that the $\mathrm{Rm_c}$ first increases sharply with decreasing $\mathrm{Pm}$, and as they further decreased $\mathrm{Pm}$, $\mathrm{Rm_c}$ decreased slowly. Haugen {\it et al.}~\cite{Haugen:PRE2004} carried out similar studies and observed that $\mathrm{Rm_c}$ decreases with increasing $\mathrm{Pm}$. In this paper, our main objective is to observe the variation of $\mathrm{Rm_c}$ with $\mathrm{Pm}$ in the two limiting cases of very high and very low $\mathrm{Pm}$.
Due to computational constraints, it is impractical to perform DNS of dynamos with very low and very high $\mathrm{Pm}$. To study these two limiting cases of $\mathrm{Pm}$, we construct a low-dimensional model by selecting five small wavenumber (large length scale) modes. We solve the magnetohydrodynamic (MHD) equations for these modes and study the bifurcations of this low-dimensional model from fluid flow to dynamo regimes of various kinds. 

Rikitake~\cite{Rikitake:MPTCPS1958} constructed a low-dimensional model for two disks dynamo using four ordinary differential equations of current and angular velocity. His model produced self-sustained dynamo and the polarity reversals of the magnetic field. Gissinger {\it et al.}~\cite{Gissinger:EPL2010} and Gissinger~\cite{Gissinger:EPJB2012} proposed a three-mode model for small $\mathrm{Pm}$ and observed field reversals due to the coupling of the dipole and the quadrupole modes, similar to the reversals of the magnetic field in geodynamo, where quadrupole mode is believed to play an important role during the reversal. In a low-dimensional model with a large number of participating modes, Donner {\it et al.}~\cite{Donner:PD2006} focused mainly on dynamo for $\mathrm{Pm}=1$. In their model described by 152 ordinary differential equations (ODEs), they observed that small wavenumber modes contain most of the magnetic and total energies. They reported constant, periodic, quasi-periodic, and chaotic dynamo states by varying the Reynolds number. 

Verma {\it et al.}~\cite {Verma:PRE2008} constructed a six-mode model containing three real velocity and three real magnetic modes and discussed the properties of pure fluid and dynamo states with and without helicity. Verma and Yadav~\cite{Verma:PP2013} constructed a three-mode model of Taylor-Green and convective dynamos, which showed crossover from supercritical dynamo transition at $\mathrm{Pm} =2$ to subcritical dynamo transition at $\mathrm{Pm} =1/2$, similar to what is observed in the numerical simulation of Taylor-Green and spherical dynamos. We note that Verma {\it et al.}~\cite {Verma:PRE2008}  as well as Verma and Yadav~\cite{Verma:PP2013} observed only the statistically steady dynamo states and no time-varying dynamo states were obtained. These variable dynamo states are an important feature of the dynamo study which has been captured by our model.

The model proposed here comprises three large-scale velocity and two large-scale magnetic modes. These modes are complex and hence, our model has ten degrees of freedom. Two of the three velocity modes are forced using Taylor-Green forcing, similar to the one used in the DNS results of Yadav {\it et al.}~\cite {Yadav:EPL2010} These five particular modes form the most dominant triadic pair interactions for the given forcing as is evident from the fact that they emerged as the most energetic modes in the DNS study of Yadav {\it et al.}~\cite {Yadav:EPL2010} Note however that these five modes were dominant in a dynamo with $\mathrm{Pm}$ of the order of unity. But the same modes may or may not be dominant for the cases when $\mathrm{Pm}$ is either very small or very large. In that case, our model may not be able to capture all the properties of the Taylor-Green dynamo for $\mathrm{Pm}$ much smaller or much larger than unity. Also, our model does not include many small-scale velocity and magnetic modes, which may also contribute to the disagreement with the DNS of Taylor-Green dynamo. 

These five modes are further decomposed onto a helical basis and only one helical component is retained. This reduction was made so that we can develop a  model with as few degrees of freedom as possible that would exhibit both dynamo transition and chaotic regimes while keeping the triadic interaction structure. In that regard, we note that our model consists of two triadic interactions. One triad involves the velocity field only and the other involves both fields (remember that there exists no triad involving the magnetic field alone).

The rest of the paper is organised as follows: we describe the MHD equations in the helical basis followed by the five-mode model in Sec.~\ref{sec:5_mode_description}. In Sec.~\ref{sec:fluid_sol}, we calculate pure fluid solutions to our model. In Sec.~\ref{sec:dyn_transition}, we emphasize on MHD solution and dynamo transition for very high and very low $\mathrm{Pm}$ limits. The various dynamo states observed for $\mathrm{Pm} =1$ are presented in Sec.~\ref{sec:time_dep_dyn}. Finally, in Sec.~\ref{sec:conclusion}, we summarize our results.

\section{Description of the five-mode model}
\label{sec:5_mode_description}

The nondimensionalised magnetohydrodynamic (MHD) equations~\cite{Verma:PR2004} are 
\begin{eqnarray}
\partial_{t}\mathbf{u}+ (\mathbf{u} \cdot \nabla) \mathbf{u} & = & -\nabla p+ (\mathbf{b} \cdot \nabla)\mathbf{b}+\nabla^{2}\mathbf{u}+\mathbf{F}, \label{eq:MHD_vel}\\
\partial_{t}\mathbf{b}+ (\mathbf{u} \cdot \nabla) \mathbf{b} & = & (\mathbf{b} \cdot \nabla) \mathbf{u}+\frac{1}{\mathrm{Pm}}\nabla^{2}\mathbf{b}, \label{eq:MHD_mag}\\
\nabla \cdot \mathbf{u} & = & 0,  \label{eq:div_u_0} \\
\nabla \cdot \mathbf{b} & = & 0,  \label{eq:div_b_0}
\end{eqnarray}
where $\mathbf{u}$ and $\mathbf{b}$ are the velocity and magnetic fields, respectively, $p$ is total pressure (thermal+magnetic), and $\mathbf{F}$ is the external force field. For nondimensionalisation, we use $L$ and $L^2/\nu$ as the length and the time scales of the system, respectively, whereas $\nu/L$ is used to scale the velocity field, and $\sqrt{\mu_0 \rho} \, \nu/L$ to scale the magnetic field. 

Following Waleffe~\cite{waleffe:PF1992} and Lessinnes {\it et al.},~\cite{Lessinnes:TCFD2009} the velocity and the magnetic fields can be expressed in a helical basis as follows
\begin{eqnarray}
\mathbf{u}(\mathbf{X}) & = & \displaystyle\sum_{\mathbf{k}}\mathbf{u}(\mathbf{k})e^{i\mathbf{k} \cdot \mathbf{X}}  \nonumber \\
& = & \displaystyle\sum_{\mathbf{k}} (u^+(\mathbf{k}) \mathbf{h^+}(\mathbf{k}) + u^-(\mathbf{k}) \mathbf{h^-}(\mathbf{k}))e^{i\mathbf{k} \cdot \mathbf{X}}, \label{eq:u_helical} \\
\mathbf{b}(\mathbf{X}) & = & \displaystyle\sum_{\mathbf{k}}\mathbf{b}(\mathbf{k})e^{i\mathbf{k} \cdot \mathbf{X}}  \nonumber \\
& = & \displaystyle\sum_{\mathbf{k}} (b^+(\mathbf{k}) \mathbf{h^+}(\mathbf{k}) + b^-(\mathbf{k}) \mathbf{h^-}(\mathbf{k}))e^{i\mathbf{k} \cdot \mathbf{X}}\,. \label{eq:b_helical}  
\end{eqnarray}

\begin{figure}[htbp]
\centering
\includegraphics[scale=0.40]{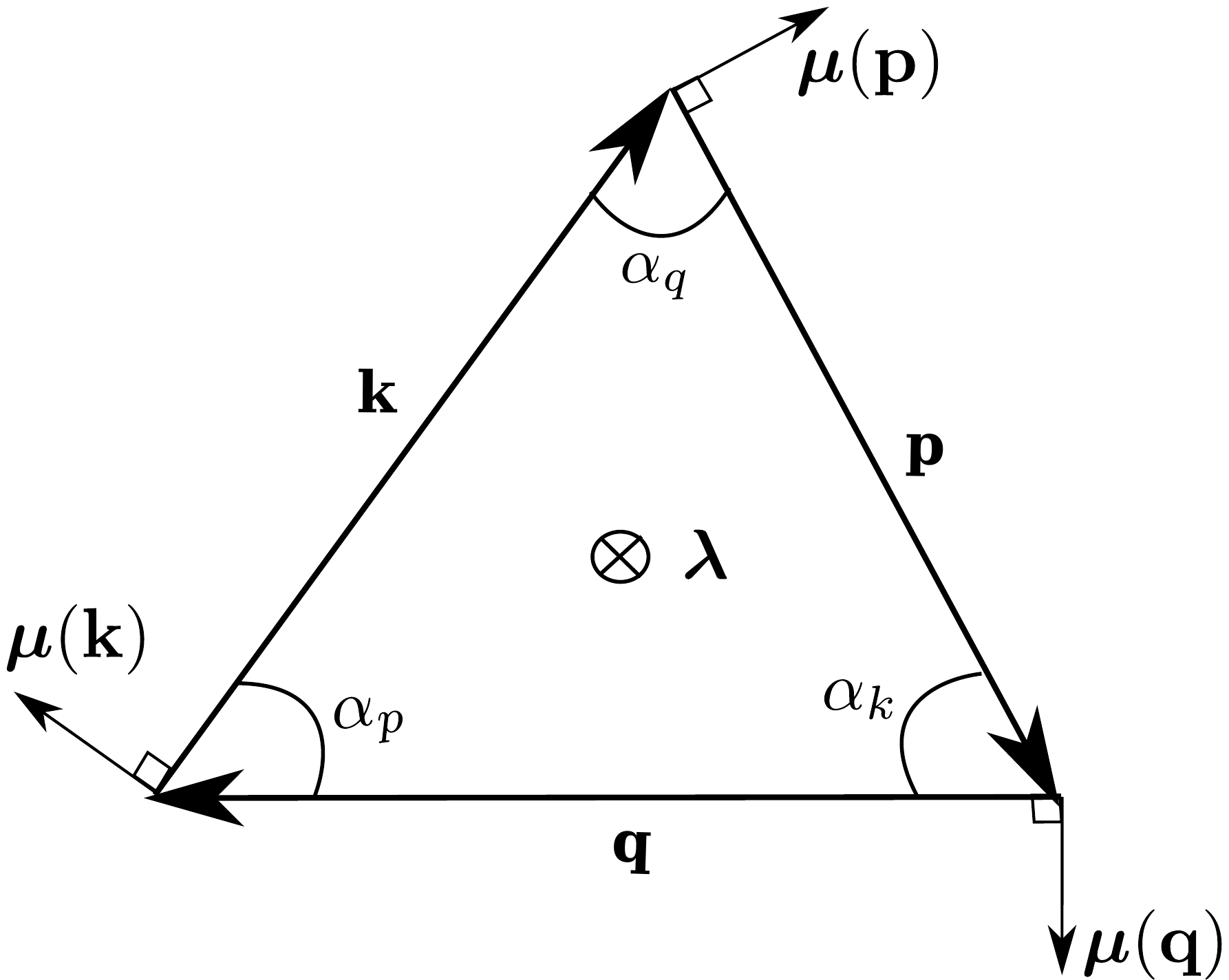}
\caption{Vector representation of triad ($\mathbf{k}$, $\mathbf{p}$, $\mathbf{q}$). This picture is inspired from Waleffe.~\cite{waleffe:PF1992}}
\label{fig:hel_triad}
\end{figure}

The $\mathbf{h^{\pm}}(\mathbf{k})$ vectors form a helical basis for the Fourier modes of the wavevector $\mathbf{k}$. They are eigenvectors of the curl operator $i \mathbf{k} \times$. In fact they can be defined up to an arbitrary rotation about $\mathbf{k}$. In practice, we follow Waleffe~\cite{waleffe:PF1992} and for each Fourier mode $\mathbf{k}$, we arbitrarily chose a vector $\boldsymbol{\nu}(\mathbf{k})$ that is orthogonal to $\mathbf{k}$. Then $\mathbf{h^\pm}(\mathbf{k})$ is defined according to
\begin{equation}
\mathbf{h}^{s_k}(\mathbf{k}) = \boldsymbol{\nu} \times \mathbf{k}/k + i s_k \boldsymbol{\nu}, \label{eq:hsk}
\end{equation}
where $k$ (as will be used from now on) is the wavenumber associated with the wavevector $\mathbf{k}$, and $s_k \pm 1$. Then, if $\mathbf{k}$, $\mathbf{p}$ and $\mathbf{q}$ are three Fourier modes forming a triad, that is $\mathbf{k} + \mathbf{p} +\mathbf{q} = 0$, we define 
\begin{eqnarray}
\boldsymbol{\lambda} &  =&  (\mathbf{k} \times \mathbf{p})/||(\mathbf{k} \times \mathbf{p})|| \nonumber \\
& & = (\mathbf{p} \times \mathbf{q})/||(\mathbf{p} \times \mathbf{q})|| = (\mathbf{q} \times \mathbf{k})/||(\mathbf{q} \times \mathbf{k})||, \label{eq:lambda}
\end{eqnarray}
a unit vector perpendicular to the plane of the triad, and 
\begin{equation}
\boldsymbol{\mu}(\mathbf{k}) =  \mathbf{k} \times \boldsymbol{\lambda}/k. \label{eq:mu}
\end{equation} 
These vectors are schematically represented in Fig.~\ref{fig:hel_triad}. Then, there exist angles $\phi_\mathbf{k}$, $\phi_\mathbf{p}$ and $\phi_\mathbf{q}$ such that 
\begin{eqnarray}
\mathbf{h}^{s_k}(\mathbf{k}) &=& e^{i s_k\, \phi_\mathbf{k}}   (\boldsymbol{\lambda}  + i s_k \boldsymbol{\mu}(\mathbf{k})),  \label{eq:hs_k} \\
\mathbf{h}^{s_p}(\mathbf{p}) &=& e^{i s_p\, \phi_\mathbf{p}}   (\boldsymbol{\lambda}  + i s_p \boldsymbol{\mu}(\mathbf{p})),  \label{eq:hs_p} \\
\mathbf{h}^{s_q}(\mathbf{q}) &=& e^{i s_q\, \phi_\mathbf{q}}   (\boldsymbol{\lambda}  + i s_q \boldsymbol{\mu}(\mathbf{q})).  \label{eq:hs_q}
\end{eqnarray}
Note that $\phi_\mathbf{k}$ is the angle of rotation around $\mathbf{k}$ needed to transform the basis $(\boldsymbol{\mu}(\mathbf{k}),\boldsymbol{\lambda})$ onto the basis $(\boldsymbol{\nu},\boldsymbol{\nu} \times \mathbf{k}/k)$. 

Substituting $\mathbf{u}$ and $\mathbf{b}$ from Eqs. (\ref{eq:u_helical}) and (\ref{eq:b_helical}) into Eqs. (\ref{eq:MHD_vel}) and (\ref{eq:MHD_mag}) and projecting on the basis $\mathbf{h}^{s_k} (s_k = \pm 1)$, the following dynamical system of equations can be obtained~\cite{Lessinnes:TCFD2009}

\begin{eqnarray}
(\partial_{t} + k^2)u^{s_k}(\mathbf{k}) & = & \frac{1}{2}\displaystyle\sum_{\mathbf{p}+\mathbf{q}+\mathbf{k}=0} \displaystyle\sum_{s_p,s_q} (s_pp - s_qq)\: g \:\nonumber \\
& & (u^{s_p}(\mathbf{p})u^{s_q} (\mathbf{q}) -b^{s_p}(\mathbf{p})b^{s_q}(\mathbf{q}))^* \nonumber\\ && +f^{s_k}(\mathbf{k}), \label{eq:MHD_u_hel}\\
(\partial_{t} +  \frac{1}{\mathrm{Pm}}k^2)b^{s_k}(\mathbf{k}) & = &  -\frac{1}{2} \displaystyle\sum_{\mathbf{p}+\mathbf{q}+\mathbf{k}=0} \displaystyle\sum_{s_p,s_q} s_kk\: g \: (u^{s_p}(\mathbf{p})b^{s_q}(\mathbf{q}) \nonumber \\ 
& &- b^{s_p}(\mathbf{p})u^{s_q}(\mathbf{q}))^*, \label{eq:MHD_b_hel} 
\end{eqnarray}
where g is defined as
\begin{eqnarray}
g(\mathbf{k}, \mathbf{p}, \mathbf{q}, s_k, s_p, s_q) & = & -\frac {1}{\mathbf{h}^{{s_k}^*}(\mathbf{k})\cdot \mathbf{h}^{s_k} (\mathbf{k})} \nonumber \\ & &  (\mathbf{h}^{{s_k}^*}(\mathbf{k}) \times \mathbf{h}^{{s_p}^*}(\mathbf{p})) \cdot \mathbf{h}^{{s_q}^*}(\mathbf{q})\,. \label{eq:g}
\end{eqnarray}

\begin{figure}[htbp]
\centering
\includegraphics[scale=1.1]{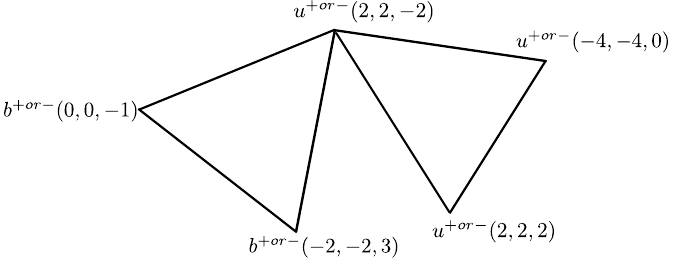}
\caption{The modes $u_1 \equiv u^{+ or -}(2,2,2)$, $u_2 \equiv u^{+ or -}(2,2,-2)$, $u_3 \equiv u^{+ or -}(-4,-4,0)$, $b_4 \equiv b^{+ or -}(0,0,-1)$, and $b_5 \equiv b^{+ or -}(-2,-2,3)$ are part of the two triads.}
\label{fig:triads}
\end{figure}

According to Eqs.~(\ref{eq:MHD_u_hel}) and (\ref{eq:MHD_b_hel}), the evolution of Fourier modes is governed by a sum of triadic interaction between modes. Three modes are in triadic interaction whenever the sum of their wavevectors vanishes. Eqs. (\ref{eq:MHD_u_hel}) and (\ref{eq:MHD_b_hel}) involve many interacting velocity and magnetic modes. In this paper, we analyse the interaction between the five modes that were found to be most energetic in the DNS study.~\cite{Yadav:EPL2010} We further focus our study on the case of the interaction between particular helical modes (recall that each Fourier mode contain two such helical modes). Specifically, we assume that all modes are zero, but for the kinetic modes of wavevectors $\mathbf{k}_1 = (2,2,2), \mathbf{k}_2 = (2,2,-2)$ and $\mathbf{k}_3 = (-4,-4,0)$ and respective helical signs $s_1= -, s_2=+$, and $s_3=-$ and for the magnetic modes $\mathbf{k}_4 = (0,0,-1)$ and $\mathbf{k}_5= (-2,-2,3)$ and respective helical signs $s_4=+$ and $s_5=-$. The state of the system is therefore specified by five complex number describing the Fourier amplitudes $u_1$, $u_2$ and $u_3$ of the three kinetic modes and the Fourier amplitudes $b_4$ and $b_5$ of the two magnetic modes. The model contains two triadic interactions (see Fig.~\ref{fig:triads}): one kinetic triad involving $u_1$, $u_2$, and $u_3$ and a magnetic triad involving $u_2$, $b_4$ and $b_5$.

Substitution of the above modes in (\ref{eq:MHD_u_hel}) and (\ref{eq:MHD_b_hel}) yields the following five ODEs
\begin{eqnarray}
\dot{u_1} & = & g_{123}(k_2 - k_3)(u_2 u_3)^* - k_1^2 u_1 + f_0, \label{eq:u_1}\\ 
\dot{u_2} & = & g_{123}(k_3 - k_1)(u_3 u_1)^* +g_{245}(k_4 - k_5)(-b_4 b_5)^* \nonumber\\ && - k_2^2 u_2 + f_0, \label{eq:u_2}\\
\dot{u_3} & = & g_{123}(k_1 - k_2)(u_1 u_2)^* - k_3^2 u_3, \label{eq:u_3}\\
\dot{b_4} & = & g_{245}(-k_4)(-u_2 b_5)^* - \dfrac{1}{\mathrm{Pm}} k_4^2 b_4, \label{eq:b_4}\\
\dot{b_5} & = & g_{245}(-k_5)(u_2 b_4)^* - \dfrac{1}{\mathrm{Pm}} k_5^2 b_5, \label{eq:b_5}
\end{eqnarray} 
where $k_1 =-3.46, k_2 =3.46, k_3 =-5.66, k_4 =1$, and $k_5 =-4.12$ are the wavenumbers (with the helicity signs) corresponding to the five modes, respectively, and $f_0 = |\mathbf{F}|$, $g_{123} = 0.47 i$, an imaginary number, and $g_{245} =-0.024 + 0.024 i$. 

In the next couple of sections, we present the analytical fluid (Sec.~\ref{sec:fluid_sol}) and MHD (Sec.~\ref{sec:dyn_transition}) solutions to the above set of equations.



\section{Fluid solution}
\label{sec:fluid_sol}

First we look for steady state solutions with zero magnetic field. Accordingly, we set $\partial_{t} \left\langle \right\rangle = 0$ and $b_4=b_5=0$ in Eqs.~(\ref{eq:u_1} - \ref{eq:b_5}) and obtain
\begin{eqnarray}
g_{123}(k_{2}-k_{3})(u_{2}u_{3})^{*}-k_{1}^{2}u_{1}+ f_{0} &=& 0, \label{eq:st_fl_eq_u1} \\
g_{123}(k_{3}-k_{1})(u_{1}u_{3})^{*}-k_{2}^{2}u_{2}+f_{0} &=& 0, \label{eq:st_fl_eq_u2} \\
g_{123}(k_{1}-k_{2})(u_{1}u_{2})^{*}-k_{3}^{2}u_{3} &=& 0. \label{eq:st_fl_eq_u3} 
\end{eqnarray}
Using Eq.~(\ref{eq:st_fl_eq_u3}), we get
\begin{eqnarray}
u_3^* = -\frac{g_{123} (k_1 - k_2) u_1 u_2}{k_3^2} \label{eq:u3_in_u1_u2}\,.
\end{eqnarray} 
Note that $g_{123}$ ($=0.47 i$) is a purely imaginary number and hence, $g_{123}^{*}=-g_{123}$. From Eqs.~(\ref{eq:st_fl_eq_u1}),~(\ref{eq:st_fl_eq_u2}), and (\ref{eq:u3_in_u1_u2}), we deduce
\begin{eqnarray}
-\left(\frac{g_{123}^2 (k_1 - k_2) (k_2 - k_3)}{k_3^2}|u_2|^2 + k_1^2\right)u_1 \nonumber \\ 
+ f_0 &=& 0\,, \label{eq:fl_u1_real} \\
-\left(\frac{g_{123}^2 (k_1 - k_2) (k_3 - k_1)}{k_3^2}|u_1|^2 + k_2^2\right)u_2 \nonumber \\ 
 + f_0 & = & 0\,. \label{eq:fl_u2_real} 
\end{eqnarray}
Because all coefficients are real, Eqs.~(\ref{eq:fl_u1_real}) and (\ref{eq:fl_u2_real}) can only be satisfied by real valued $u_1$ and $u_2$. Then Eq.~(\ref{eq:u3_in_u1_u2}) implies that for a fluid stationary state, $u_3$ is an imaginary number. As a result, we have $u_{1}=u_{1}^{*}$, $u_{2}=u_{2}^{*}$, and $u_{3}=-u_{3}^{*}$. 

We next proceed to get an analytical solution for the fluid modes $u_{1}, u_{2}$, and $u_{3}$. It is not possible to obtain the relationship between the fluid modes and the forcing $f_0$ in closed form. Hence, we attempt to get approximate solutions. Towards this end, we concentrate on two extreme asymptotic limits of $f_0 \ll 1$ and $f_0 \gg 1$. Before that, we note that the fluid modes $u_1$ and $u_2$ have the same effective forcing for the fluid solutions due to the absence of the magnetic field. For sake of simplicity we restrict ourselves to the case $f_0 > 0$. The case of $f_0 < 0$ has the forcing with the direction of rotation reversed. The trivial fluid solution of the system is one where the fluid modes have the same direction of rotation as the forcing and hence $u_1 > 0$ and $u_2 > 0$. We note that there are other non-trivial fluid solutions possible. However, the dynamo transition -- which is the main focus of the current paper -- for the other fluid solutions appear at much larger values of forcing $f_0$ and $\mathrm{Rm}$ as compared to the trivial fluid solution. Hence, we concentrate in this paper only on the trivial fluid solution.

We first start with the case of $f_0 \ll 1$. For balance, we require the first term in both Eqs.~(\ref{eq:fl_u1_real}) and (\ref{eq:fl_u2_real}) to be $\ll 1$. This is only possible when $u_{1} \ll 1$ and $u_{2} \ll 1$. To understand this observation, we note that both $k_1^2$ and $k_2^2$ are positive numbers of $\mathcal{O}(1)$. The quantity $\displaystyle \frac{g_{123}^2 (k_1 - k_2) (k_2 - k_3)}{k_3^2} >0$ (note that $g_{123}$ is a purely imaginary number) and hence, the effective coefficient of $u_1$  in Eq.~(\ref{eq:fl_u1_real}) is at least $\mathcal{O}(1)$ quantity. Hence, balance in Eq.~(\ref{eq:fl_u1_real}) demands that $u_1 \ll 1$. Substitution of this information in Eq.~(\ref{eq:fl_u2_real}) then leads to the conclusion that $u_2 \ll 1$ as well. Hence for small forcing, i.e., $f_0 \ll 1$, we have $ u_1 \approx u_2 \approx f_0$.

We now focus on the other limit of $f_0 \gg 1$. From Eq.~(\ref{eq:fl_u2_real}), we can note that both $u_{1} \gg 1$ and $u_{2} \gg 1$ are not possible because the term $|u_{1}|^2 u_2$ will dominate $u_2$ in Eq.~(\ref{eq:fl_u2_real}) but will not balance $f_0$ since the coefficient of $|u_{1}|^2 u_2$ is positive, the same sign as that of $f_0$. There are two possibilities now for balance in Eq.~(\ref{eq:fl_u2_real}); a) $u_2$ balances $f_0$ while $|u_{1}|^2 u_2 \ll u_2$, implying $|u_{1}|^2 \ll 1$ and $u_2 \approx \mathcal{O}(f_0)$; or b) $|u_{1}|^2 u_2 \approx \mathcal{O}(f_0)$ and $u_2$ balances both of them, implying $u_1 \approx \mathcal{O}(1)$ while $u_2 \approx \mathcal{O}(f_0)$.  However, we can show that the scenario (b) is not possible by considering balance in  Eq.~(\ref{eq:fl_u1_real}). If $u_1 \approx \mathcal{O}(1)$ and $u_2 \approx \mathcal{O}(f_0)$ with $f_0 \gg 1$,  the term $|u_2|^2u_1 \gg f_0$ remains unbalanced in Eq.~(\ref{eq:fl_u1_real}). However, if $u_1 \ll 1$, the two dominant terms in Eq.~(\ref{eq:fl_u1_real}) are $|u_2|^2 u_1$ and $f_0$ which can balance if $u_2 \approx \mathcal{O}(f_0)$ while $u_1 \approx \mathcal{O}(1/f_0)$, which is consistent with the scenario (a) for balance in  Eq.~(\ref{eq:fl_u2_real}). Hence, we conclude that the trivial fluid solution for $f_0 \gg 1$ has  $u_2 \approx \mathcal{O}(f_0)$ and $u_1 \approx \mathcal{O}(1/f_0)$.

For both the limiting cases of very large and very small forcing $f_0$, we have $u_1 \ll 1$ and hence, we can make an approximation that $u_1 \ll 1$ for the entire range of forcing. With this approximation, we can solve for $u_2$ from Eq.~(\ref{eq:fl_u2_real}) as 
\begin{eqnarray}
u_2 \approx \frac{f_0}{k_2^2} = \frac{f_0}{12}\,. 
\label{eq:sol_fl_u2}
\end{eqnarray}
Substituting this solution for $u_2$ in Eq.~(\ref{eq:fl_u1_real}), we get 
\begin{eqnarray}
u_1 \approx \frac{f_0}{k_1^2 + \frac{g_{123}^2 (k_1 - k_2) (k_2 - k_3) f_0^2}{k_1^4 k_3^2}} = \frac{f_0}{12 + 0.0031 f_0^2}\,. 
 \label{eq:sol_fl_u1}
\end{eqnarray} 
Finally, using Eq.~(\ref{eq:u3_in_u1_u2}), we estimate $u_3$ as 
\begin{eqnarray}
u_3 &\approx& \frac{g_{123} (k_1 - k_2) f_0^2}{k_2^2 k_3^2\left(k_1^2 + \frac{g_{123}^2 (k_1 - k_2) (k_2 - k_3) f_0^2}{k_1^4 k_3^2}\right)}  \nonumber \\
&=& \frac{-3.25 i f_0^2}{4608 + 1.17 f_0^2}. 
 \label{eq:sol_fl_u3}
\end{eqnarray} 
Note that for very small $f_0$, $u_1 \approx u_2$, whereas for very large $f_0$, $u_1$ is very small as compared to $u_2$.

In the next section, we analytically obtain estimates for the constant  MHD solution and use it to calculate the critical magnetic Reynolds number for dynamo transition in the two limiting cases of very small and very large magnetic Prandtl numbers.

\section{MHD solution and dynamo transition}
\label{sec:dyn_transition}
In this section, we first focus on the analytical solutions to the three velocity and the two magnetic modes for the MHD state. We will then use these to relate the velocity modes in terms of $\mathrm{Pm}$ just above the dynamo onset. Later, we emphasize on the two limiting cases of $\mathrm{Pm}$ to calculate the critical magnetic Reynolds number, $\mathrm{Rm_c}$, corresponding to the dynamo action for both the cases. 

For the steady state (constant) MHD solution, we set $\partial_{t} \left\langle \right\rangle = 0$ in Eqs.~(\ref{eq:u_1} - \ref{eq:b_5}). For $b_{4} \neq 0$ and $b_{5} \neq 0$ in Eqs.~(\ref{eq:b_4}) and (\ref{eq:b_5}), we require 
\begin{eqnarray}
|u_2|=\dfrac{\sqrt{-k_{4}k_{5}}}{\mathrm{Pm}|g_{245}|}. \label{eq:gen_res_u2_1}
\end{eqnarray}
We note that the steady state equations corresponding to the $u_1$ and $u_3$ velocity modes remain the same as for the fluid solution (Eqs.~(\ref{eq:st_fl_eq_u1}) and (\ref{eq:st_fl_eq_u3})).  Hence, we have 
\begin{eqnarray}
u_{3}=\dfrac{g_{123}(k_{1}-k_{2})(u_{1}u_{2})^{*}}{k_{3}^{2}} \label{eq:u_3_a}
\end{eqnarray}
which upon substitution in Eq.~(\ref{eq:st_fl_eq_u1}) gives us
\begin{eqnarray}
\left(\dfrac{|g_{123}|^{2}(k_{1}-k_{2})(k_{2}-k_{3})|u_{2}|^{2}}{k_{3}^{2}} - k_{1}^{2}\right)u_{1} \nonumber \\ 
+ f_{0} &=& 0\,. \label{eq:eqn_in_u1}
\end{eqnarray}
Substituting $|u_2|$ from Eq.~(\ref{eq:gen_res_u2_1}) in the above, we can solve for $u_1$ in terms of $f_0$ as
\begin{widetext}
\begin{eqnarray}
u_{1} = \left(\dfrac{|g_{245}|^2k_{3}^{2}\mathrm{Pm}^2 }{|g_{123}|^{2}(k_{1}-k_{2})(k_{2}-k_{3})k_4\,k_5+ |g_{245}|^2k_{1}^{2}k_{3}^{2}\mathrm{Pm}^2}\right)  f_{0} \,. \label{eq:mhd_u1}
\end{eqnarray}
\end{widetext}
From the above, we can conclude that the velocity mode $u_1$ for the MHD solution is real as well. We next need to determine the nature of the velocity mode $u_2$. Towards this end, we solve for $b_5$ from Eq.~(\ref{eq:b_5}) and substitute in Eq.~(\ref{eq:u_2}) (with $\partial_{t} \left\langle \right\rangle = 0$) to get 
\begin{widetext}
\begin{eqnarray}
\left(\dfrac{|g_{123}|^{2}(k_{1}-k_{2})(k_{3}-k_{1}) |u_{1}|^{2}}{k_{3}^{2}}+ \dfrac{|g_{245}|^2(k_4 - k_5) \mathrm{Pm}|b_4|^2}{k_5} - k_{2}^{2}\right)u_{2}  + f_{0} =0 \label{eq:eqn_in_u2}
\end{eqnarray}
\end{widetext}

Since $f_{0}$ and the coefficient of $u_{2}$ in the above equation are real numbers, we conclude that $u_{2}$ is a real number for the MHD state as well. Also, from Eq.~(\ref{eq:u_3_a}), it is evident that $u_3$ is a purely imaginary number. Therefore, the velocity mode $u_2$ and $u_3$ for the MHD state are
\begin{eqnarray}
u_2=\dfrac{\sqrt{-k_{4}k_{5}}}{\mathrm{Pm}|g_{245}|} \label{eq:mhd_u2}
\end{eqnarray}
and
\begin{widetext}
\begin{eqnarray}
u_{3}= \left(\dfrac{g_{123}(k_{1}-k_{2})\sqrt{-k_{4}k_{5}}|g_{245}|\mathrm{Pm}}{|g_{123}|^{2}(k_{1}-k_{2})(k_{2}-k_{3})k_4\,k_5+ |g_{245}|^2k_{1}^{2}k_{3}^{2}\mathrm{Pm}^2}\right) f_{0}\,. \label{eq:mhd_u3}
\end{eqnarray}
\end{widetext}
To get the magnitude of the magnetic mode $b_4$ in terms of $f_0$ and $\mathrm{Pm}$, we can substitute Eqs.~(\ref{eq:mhd_u1}) and (\ref{eq:mhd_u2}) in Eq.~(\ref{eq:eqn_in_u2}) and solve for $|b_4|$. Finally, $b_5$ can be obtained by substituting for $b_4$  and $u_2$ in Eq.~(\ref{eq:b_5}) with $\partial_{t} b_5 = 0$. The resulting expressions become fairly long and are not reported here. Instead, we can solve for the magnitude of the magnetic modes in terms of the velocity modes using Eq.~(\ref{eq:u_2}), (\ref{eq:b_4}), and (\ref{eq:b_5}) as
\begin{eqnarray}
|b_4| &=& \sqrt{\dfrac{[g_{123}(k_3 - k_1) u_1 u_3 + k_2^2 u_2 - f_0]k_5}{|g_{245}|^2 (k_4 - k_5) u_2 \mathrm{Pm}}} \label{eq:sol_b4}, \\
|b_5| &=& \sqrt{\dfrac{[g_{123}(k_3 - k_1) u_1 u_3 + k_2^2 u_2 - f_0] u_2 \mathrm{Pm}}{(k_4 - k_5) k_5}}. \label{eq:sol_b5}
\end{eqnarray}
Since $g_{245}=0.024(-1+ i)$ is neither purely real nor purely imaginary, we observe from Eqs.~(\ref{eq:b_4}) and (\ref{eq:b_5}) that both the magnetic modes $b_4$ and $b_5$ are complex numbers. The magnetic modes in terms of $f_0$ and $\mathrm{Pm}$ can be obtained by substituting for $u_1, u_2$, and $u_3$ from Eqs.~(\ref{eq:mhd_u1}), (\ref{eq:mhd_u2}), and (\ref{eq:mhd_u3}), respectively in Eqs.~(\ref{eq:sol_b4}) and (\ref{eq:sol_b5}). This gives us a one-parameter family for the steady dynamo solution in terms of $f_0$ for a given $\mathrm{Pm}$. From Eqs.~(\ref{eq:sol_b4}) and (\ref{eq:sol_b5}), we observe that the magnetic modes follow $|b_4|/|b_5| = \sqrt{-k_5/k_4} \approx 2$ for our model irrespective of $\mathrm{Pm}$ and $f_0$. 

We next focus on the critical magnetic Reynolds number $\mathrm{Rm_c}$ and critical forcing $f_c$ for the onset of dynamo action. Near the dynamo onset, the magnetic modes are small and hence we can assume the term containing $|b_{4}|^2$ in Eq.~(\ref{eq:eqn_in_u2}) to be negligible in comparison with the other terms. With this approximation and using the fact that $u_1$ and $u_2$ are real numbers, elimination of $f_0$ from Eq.~(\ref{eq:eqn_in_u1}) and (\ref{eq:eqn_in_u2}) leads to
\begin{widetext}
\begin{eqnarray}
\left(\dfrac{|g_{123}|^{2}(k_{1}-k_{2})(k_{3}-k_{1})u_{2}}{k_{3}^{2}}\right)u_{1}^{2} - \left(\dfrac{|g_{123}|^{2}(k_{1}-k_{2})(k_{2}-k_{3})u_{2}^{2}}{k_{3}^{2}} - k_{1}^{2}\right)u_{1} -k_{2}^{2}u_{2} = 0. \label{eq:quad_in_u1}
\end{eqnarray}
\end{widetext}
Substituting for $u_2$ from Eq.~(\ref{eq:mhd_u2}), the above equation reduces to a quadratic equation in $u_{1}$ which can be solved easily. Finally, putting $u_{1}$ and $u_{2}$ in Eq.~(\ref{eq:u_3_a}), we can get the solution for $u_{3}$. The final analytical expression for the velocity modes at the onset of the dynamo is obtained as
\begin{widetext}
\begin{eqnarray}
u_{1} &=& \dfrac{2.37}{\mathrm{Pm}}(-52.79 - 0.39 \: \mathrm{Pm}^{2} + \sqrt{2786.91 +62.54\: \mathrm{Pm}^{2} +0.16 \: \mathrm{Pm}^{4}}) \label{eq:gen_res_u1} \\
u_{2} &=& \dfrac{60.08}{\mathrm{Pm}} \label{eq:gen_res_u2} \\
u_{3} &=& \dfrac{-14.51 \: i}{\mathrm{Pm}^{2}}(-52.79-0.39 \: \mathrm{Pm}^{2} + \sqrt{2786.91 + 62.54\: \mathrm{Pm}^{2} +0.16 \: \mathrm{Pm}^{4}}). \label{eq:gen_res_u3}
\end{eqnarray}
\end{widetext}
In the above, we have substituted for the numerical values of the various constants and rounded off to the second decimal place. However, to get the limiting values below, we will first obtain the limits and then round off to the second decimal place. We note that there are two roots of the quadratic Eq.~(\ref{eq:quad_in_u1}). However, the other root for $u_1$ is negative and hence we do not consider them for the reason discussed in Sec.~\ref{sec:fluid_sol}. It has been observed in the literature that the $\mathrm{Rm_c}$ saturates for the limiting cases of very low and very high $\mathrm{Pm}$.~\cite{Ponty:PRL2005,Nigro:APJ2011} In order to ascertain this behavior in our low-dimensional model, we look for the solutions to the velocity modes in the two limiting cases of very low and very high $\mathrm{Pm}$, which are as follows \\
{For low $\mathrm{Pm}$ ($\mathrm{Pm} \rightarrow 0$), we get} 
\begin{eqnarray}
u_{1} & \approx & 0.46 \, \mathrm{Pm}, \label{eq:small_res_u1} \\ 
u_{2} & \approx & \dfrac{60.08}{\mathrm{Pm}}, \label{eq:gen_small_u2} \\
u_{3} & \approx & -2.79\:i. \label{eq:small_res_u3}
\end{eqnarray}

Hence, for our nondimensionalised MHD equations, 
\begin{eqnarray}
\mathrm{Rm}_c=\left(\sqrt{|u_{1}|^{2}+|u_{2}|^{2}+|u_{3}|^{2}}\right)\mathrm{Pm} \approx 60.08. \label{eq:small_Rmc}
\end{eqnarray}
{The velocity modes for high $\mathrm{Pm}$ ($\mathrm{Pm} \rightarrow \infty$) are} 
\begin{eqnarray}
u_{1} & \approx & u_{2} = \dfrac{60.08}{\mathrm{Pm}} , \label{eq:large_res_u1_u2} \\ 
u_{3} & \approx & \dfrac{-368.34 \:i}{\mathrm{Pm}^{2}}. \label{eq:large_res_u3}
\end{eqnarray}

Hence, 
\begin{eqnarray}
\mathrm{Rm}_c \approx 84.90. \label{eq:large_Rmc}
\end{eqnarray}

\begin{figure}[htbp]
\centering
\includegraphics[scale=0.48]{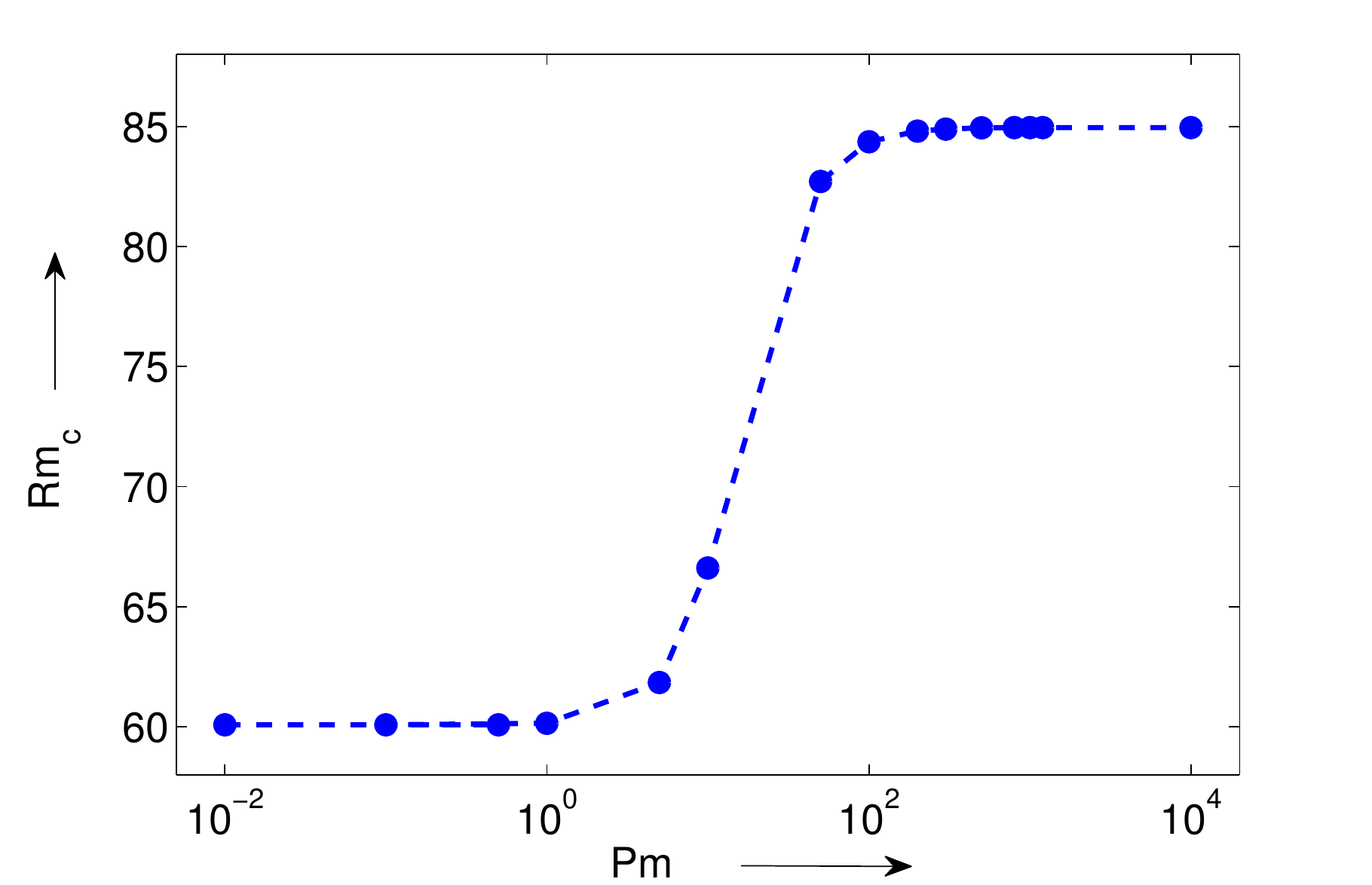}
\caption{Plot of critical magnetic Reynolds number for dynamo ($\mathrm{Rm_c}$) with magnetic Prandtl number ($\mathrm{Pm}$), which shows that the $\mathrm{Rm_c}$ saturates to $60.08$ for very low $\mathrm{Pm}$ and to $84.90$ for very high $\mathrm{Pm}$.}
\label{fig:Rmc_vs_Pm}
\end{figure}

Therefore, our model shows that $\mathrm{Rm_c}$ saturates to a constant value in both the limits of very high and very low $\mathrm{Pm}$ (depicted in Fig.~\ref{fig:Rmc_vs_Pm}). Ponty {\it et al.}~\cite{Ponty:PRL2005}, Nigro and Veltri~\cite{Nigro:APJ2011} have also reported saturation of $\mathrm{Rm}_c$ for small $\mathrm{Pm}$. The limiting value of $\mathrm{Rm}_c$ for high $\mathrm{Pm}$ is higher than that of low $\mathrm{Pm}$. However, Ponty {\it et al}.~\cite{Ponty:PRL2005} observed that $\mathrm{Rm}_c$ increases with decreasing $\mathrm{Pm}$. Similar results have been reported by Haugen {\it et al}.~\cite{Haugen:PRE2004}. But in our case, we observe $\mathrm{Rm}_c$ to be small for smaller $\mathrm{Pm}$. For very small $\mathrm{Pm}$, the system would become more turbulent, and in that case small-scale modes would play a crucial role in the dynamo process. Those small-scale velocity and magnetic modes are absent in our model, which may be the reason why we observe $\mathrm{Rm}_c$ vs $\mathrm{Pm}$ trend not in agreement with that of DNS, where the contributions of small-scale modes are accounted for. On the other hand, Nigro and Veltri~\cite{Nigro:APJ2011} used a shell model to study the dynamo transition for very small and very large $\mathrm{Pm}$. They observed that $\mathrm{Rm}_c$  for $\mathrm{Pm} \gg 1$ is larger than that for $\mathrm{Pm} \ll 1$, similar to our findings. 

\begin{figure}[htbp]
\centering
\includegraphics[scale=0.35]{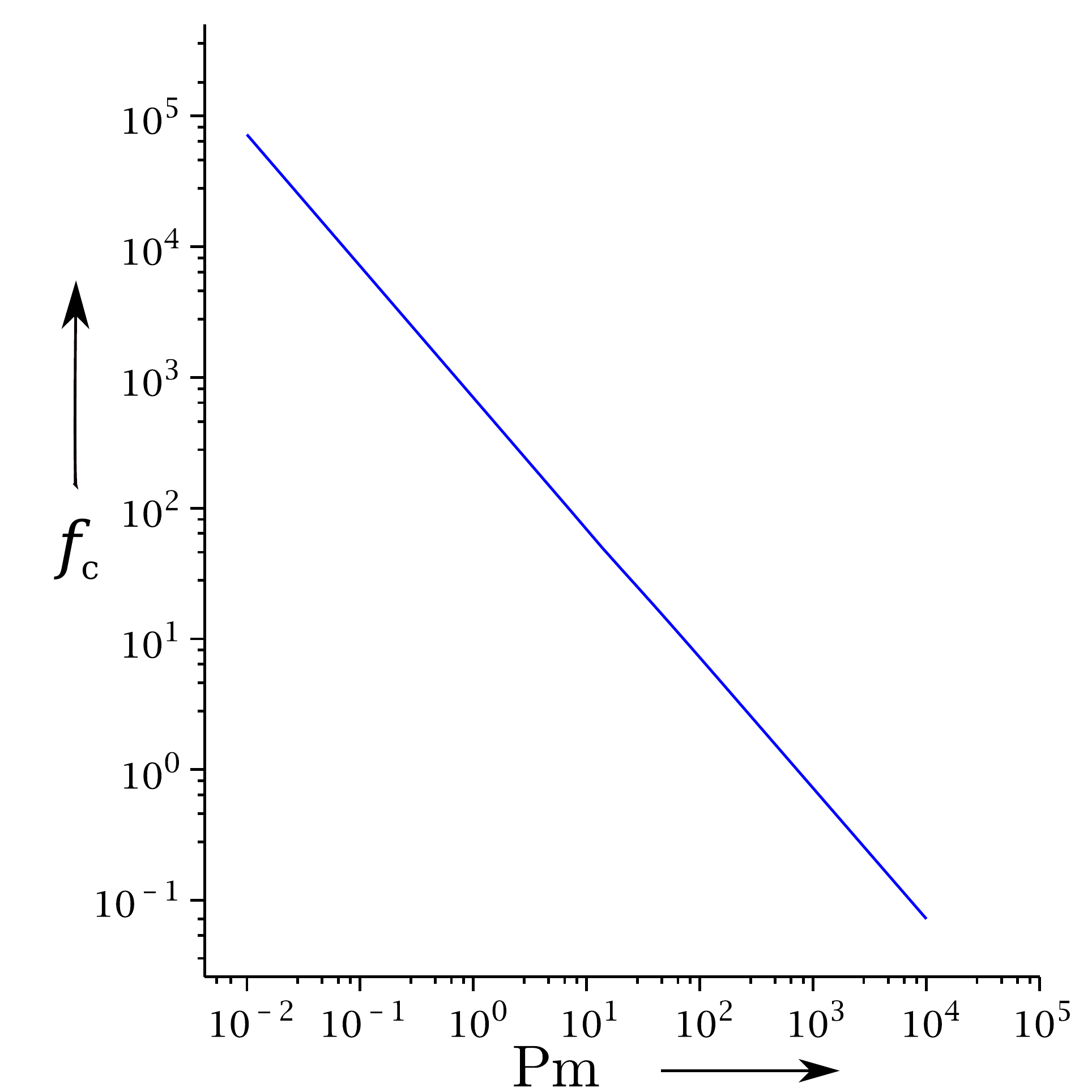}
\caption{Plot showing the variation of critical forcing for dynamo ($f_c$) with $\mathrm{Pm}$.}
\label{fig:critical_force}
\end{figure}

We next investigate the dependence of the critical forcing for dynamo ($f_c$) on $\mathrm{Pm}$.  For this purpose, we equate the velocity mode $u_2$ obtained for the fluid and the MHD solutions to get $f_c$ in terms of $\mathrm{Pm}$ as
\begin{eqnarray}
f_c = k_2^2 \dfrac{\sqrt{-k_{4}k_{5}}}{\mathrm{Pm}|g_{245}|} \approx \dfrac{720.96}{\mathrm{Pm}}\,. \label{eq:Fc}
\end{eqnarray}
This behavior has been shown in Fig.~\ref{fig:critical_force}, where we show the variation of $f_c$ with $\mathrm{Pm}$ on the  log-log scale. The critical forcing decreases continuously with increasing $\mathrm{Pm}$ and hence our model indicates that the dynamo transition becomes easier as we increase the $\mathrm{Pm}$, in accordance with the observations from DNS~\cite{Yadav:EPL2010,Yadav:PRE2012}. Hence, even though our model shows that a decrease in  $\mathrm{Pm}$ decreases the $\mathrm{Rm_c}$, it still captures the fact that it is difficult to initiate dynamo for low  $\mathrm{Pm}$ as the critical forcing amplitude $f_c$ increases sharply. 

We finally focus on the scaling of the magnetic energy $E_b$ with the forcing amplitude $f_0$ and the magnetic Reynolds number $\mathrm{Rm}$ in the two limiting cases of very large and very small $\mathrm{Pm}$.  The non-dimensional magnetic energy for our model is given by 
\begin{equation}
E_b= \frac{|b_4|^2 + |b_5|^2}{2} = \dfrac{[f_0-g_{123}(k_3 - k_1) u_1 u_3 - k_2^2 u_2]}{2\,|g_{245}|^2 u_2 \mathrm{Pm}}
\end{equation}
using Eqs.~(\ref{eq:sol_b4})  and (\ref{eq:sol_b5}).
In our model $u_3$ is always very small compared to $u_1$ and $u_2$, so we neglect the term containing $u_3$. Using Eqs.~(\ref{eq:mhd_u2}) and (\ref{eq:Fc}), we get
\begin{equation}
E_b \approx \dfrac{f_0 - f_c}{2 |g_{245}| \sqrt{-k_{4}k_{5}}} \approx 7.3\left(f_0-f_c\right)\,. 
\label{eq:EB}
\end{equation}
It can be observed from Eq.~(\ref{eq:EB}) that the nature of the bifurcation to the dynamo state is always supercritical irrespective of Pm, i.e., we do not get magnetic modes for $f_0<  f_c$. However, DNS results for low $ \mathrm{Pm}$ have shown the bifurcation to be subcritical.~\cite{Yadav:PRE2012} Krstulovic {\it et al.}~\cite{Krstulovic:PRE2011} reported the dynamo transition through a supercritical bifurcation for large $\mathrm{Pm}$, whereas for small $\mathrm{Pm}$ it was through a subcritical one. They attributed the subcritical dynamo transition to the presence of a hydrodynamic instability which affects the growing magnetic modes. For dynamo simulations in a rotating spherical shell, Morin and Dormy~\cite{Morin:IJMPB2009} have also observed a supercritical dynamo transition for a large $\mathrm{Pm}$, and a subcritical transition for a smaller $\mathrm{Pm}$. Our low-dimensional model does not capture this feature of the dynamo. This is probably because several other modes which become important for low $\mathrm{Pm}$ have been neglected in our present model. More refined low dimensional models required to capture subcritical dynamo transition would be attempted in our future work. 

\begin{figure}[htbp]
\centering
\includegraphics[scale=0.42]{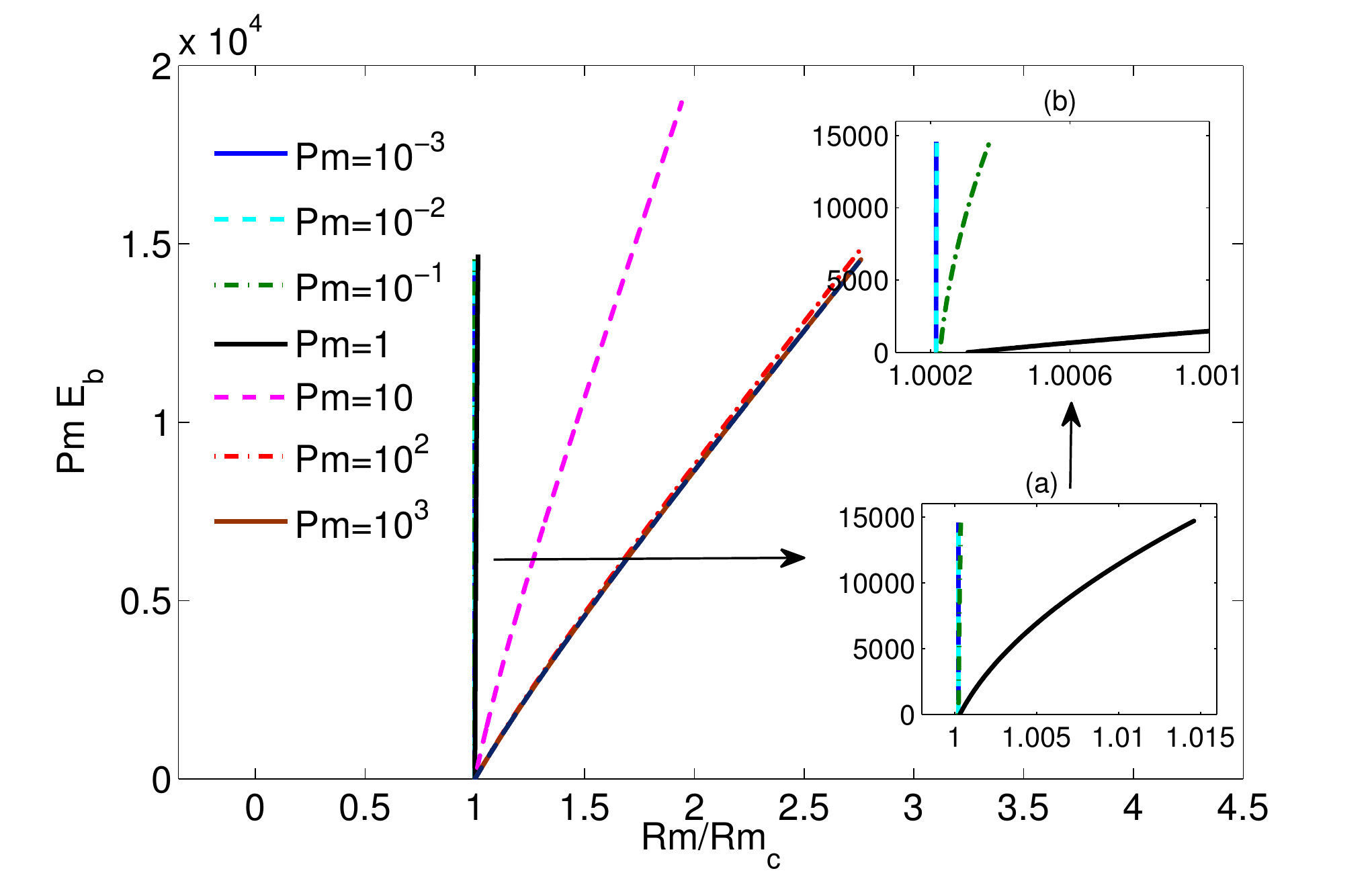}
\caption{Plot showing the variation of the magnetic energy ($E_b$) multiplied by $\mathrm{Pm}$ with reduced magnetic Reynolds number ($\mathrm{Rm}/\mathrm{Rm}_c$). The zoomed view of the plots for $\mathrm{Pm} =10^{-3}, 10^{-2}, 10^{-1}$, and $1$ is shown in the subplot (a). In the subplot (b), further zoomed view of the subplot (a) is shown.}
\label{fig:mag_energy_Rm}
\end{figure}

To estimate the magnetic energy $E_b$ in terms of $\mathrm{Rm}$ near the dynamo onset for low and high $\mathrm{Pm}$, we notice that $\mathrm{Rm}$ near the dynamo onset will approximately be determined by the fluid solution, i.e., Eqs.~(\ref{eq:sol_fl_u2} - \ref{eq:sol_fl_u3}). Also $f_0=\mathcal{O}(f_c)$ near the dynamo onset and hence, $f_0 \gg 1$ for the low $\mathrm{Pm}$ ($\mathrm{Pm}\ll 1$) case while $f_0 \ll 1$ for the high $\mathrm{Pm}$ case [see Fig.~\ref{fig:critical_force}]. Hence, for $\mathrm{Pm} \rightarrow 0$, we have $u_1 \ll 1$ with $\displaystyle u_2 \approx f_0/12$ and $u_3 \approx 2.78$. Clearly $u_2 \gg u_3 \gg u_1$  and hence, we have 
$$ \mathrm{Rm}=\left(\sqrt{|u_{1}|^{2}+|u_{2}|^{2}+|u_{3}|^{2}}\right)\mathrm{Pm} \approx \frac{\mathrm{Pm}}{12} f_0.$$
 The above can be solved for $f_0$ in terms of $\mathrm{Rm}$ and substituted in Eq.~(\ref{eq:EB}) to get 
(for $\mathrm{Pm} \rightarrow 0$) 
\begin{eqnarray}
E_b & \approx & \frac{88.0}{\mathrm{Pm}} (\mathrm{Rm} - \mathrm{Rm}_c), 
\end{eqnarray}
Similarly for $\mathrm{Pm} \rightarrow \infty$, we have $f_0 \rightarrow 0$ and accordingly, have $u_1 \approx u_2 \approx f_0/12$ while $u_3 \ll 1$. Hence, we get the relationship between $\mathrm{Rm}$ and $f_0$ as  $\displaystyle \mathrm{Rm} \approx \frac{\sqrt{2}\mathrm{Pm}}{12} f_0$  which finally results in 
\begin{eqnarray}
E_b & \approx & \frac{62.0}{\mathrm{Pm}} (\mathrm{Rm} - \mathrm{Rm}_c), 
\end{eqnarray}
Taking into account the presence of $\mathrm{Pm}$ in the denominator of the above relations between the magnetic energy $E_b$ and the magnetic Reynolds number $\mathrm{Rm}$, it is evident that the growth of $E_b$ with $\mathrm{Rm}$ for small $\mathrm{Pm}$ is much faster than that for large $\mathrm{Pm}$. In Fig.~\ref{fig:mag_energy_Rm}, we show the variation of the magnetic energy multiplied by $\mathrm{Pm}$ with the reduced magnetic Reynolds number ($\mathrm{Rm}/\mathrm{Rm}_c$). The plots are for $\mathrm{Pm} =10^{-3}, 10^{-2}, 10^{-1}, 1, 10^{1}, 10^{2}$, and $10^{3}$. As the magnetic Prandtl number decreases, slope of $\mathrm{Pm}E_b$ increases, and the magnetic modes increase sharply with an increase in forcing beyond the dynamo transition for low $\mathrm{Pm}$. This is consistent with the findings of P{\'e}tr{\'e}lis and Fauve.~\cite{Petrelis:EPJB2001} As observed earlier, we always get a supercritical bifurcation only.

After discussing the features of MHD solution and dynamo transition for a wide range of magnetic Prandtl numbers, in the next section, we concentrate on various time-dependent states in the dynamo for $\mathrm{Pm}=1$.

\section{Time-dependent dynamo solutions for $\mathrm{Pm} =1$}
\label{sec:time_dep_dyn}
  
\begin{figure}[htbp]
\centering
\includegraphics[scale=0.45]{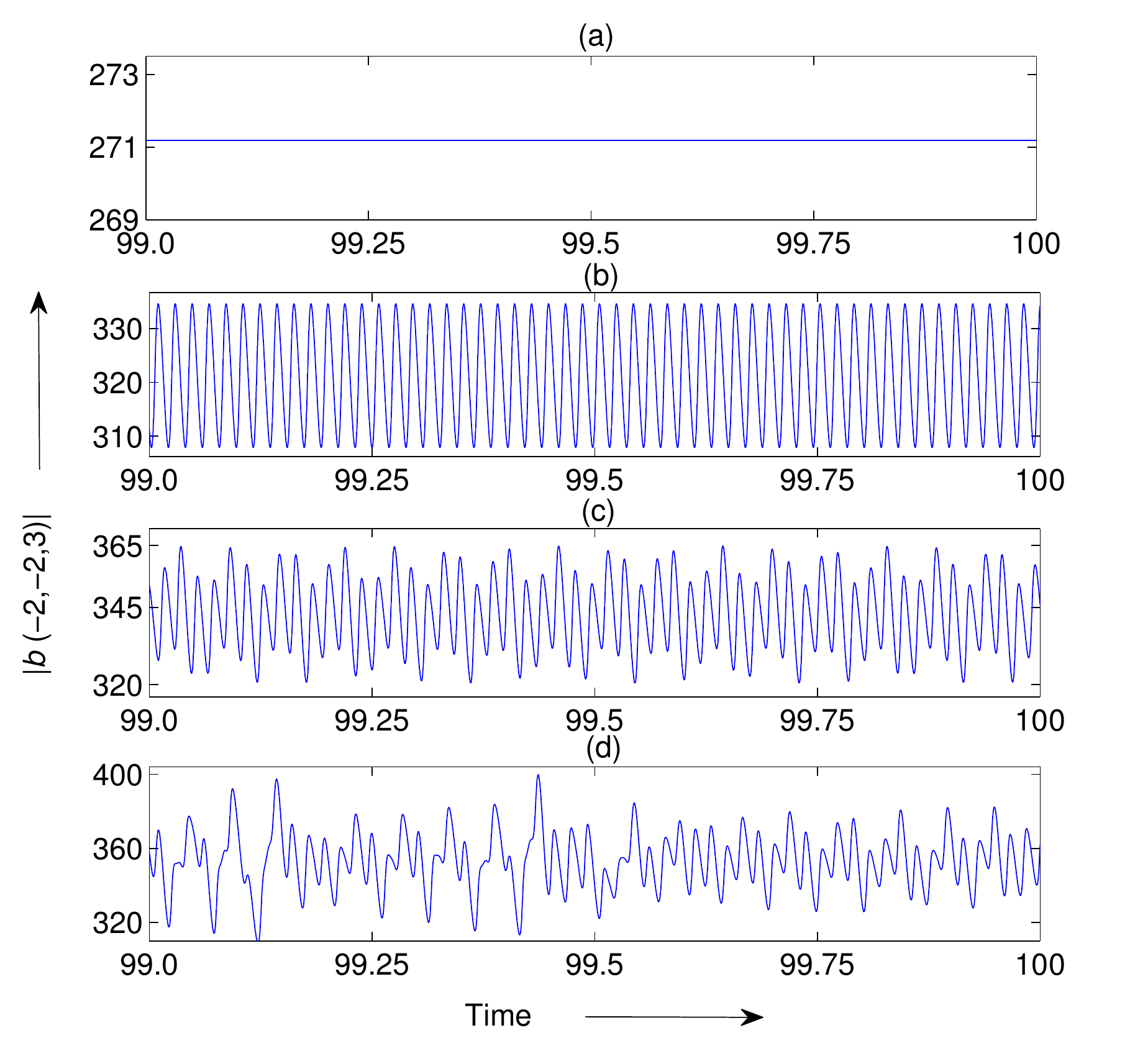}
\caption{For $\mathrm{Pm} = 1$: Time-evolution of magnetic mode $|b(-2,-2,3)|$ showing (a) fixed point solution for $f_0 = 25000$, (b) periodic solution for $f_0 = 35000$, (c) quasi-periodic solution for $f_0 = 40000$, and (d) chaotic solution for $f_0 = 42500$. }
\label{fig:time_series}
\end{figure}

\begin{figure}[htbp]
\centering
\includegraphics[scale=0.45]{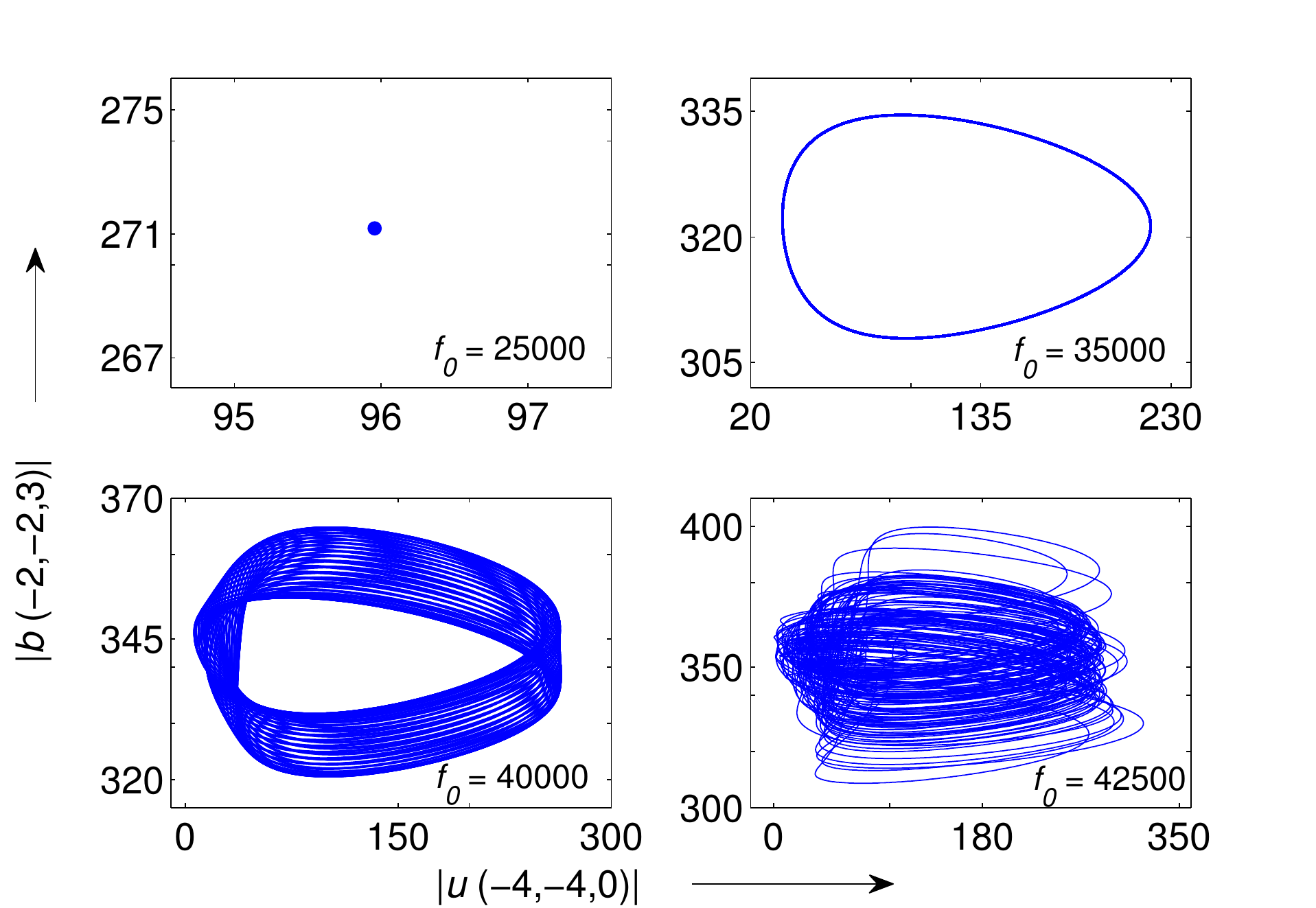}
\caption{For $\mathrm{Pm} = 1$: Phase space projections on the plane of $|b(-2,-2,3)|$ and $|u(-4,-4,0)|$ showing fixed point magnetic field for $f_0 = 25000$, periodic for $f_0 = 35000$, quasi-periodic for $f_0 = 40000$, and chaotic for $f_0 = 42500$. }
\label{fig:phase_space}
\end{figure}

In this section, we look for different time-dependent solutions to the velocity and magnetic modes after the dynamo transition for $\mathrm{Pm}=1$. The critical forcing corresponding to the dynamo action for $\mathrm{Pm}=1$ is $f_c \approx 720$ (numerical). As we further increase the forcing amplitude, we observe stationary, periodic, quasi-periodic, and chaotic dynamo states. The time-series of various dynamo states is shown in Fig.~\ref{fig:time_series}. Also, the phase space projection on ($|u_3|, |b_5|$) plane is shown in Fig.~\ref{fig:phase_space}. 

\begin{figure}[htbp]
\centering
\includegraphics[scale=0.11]{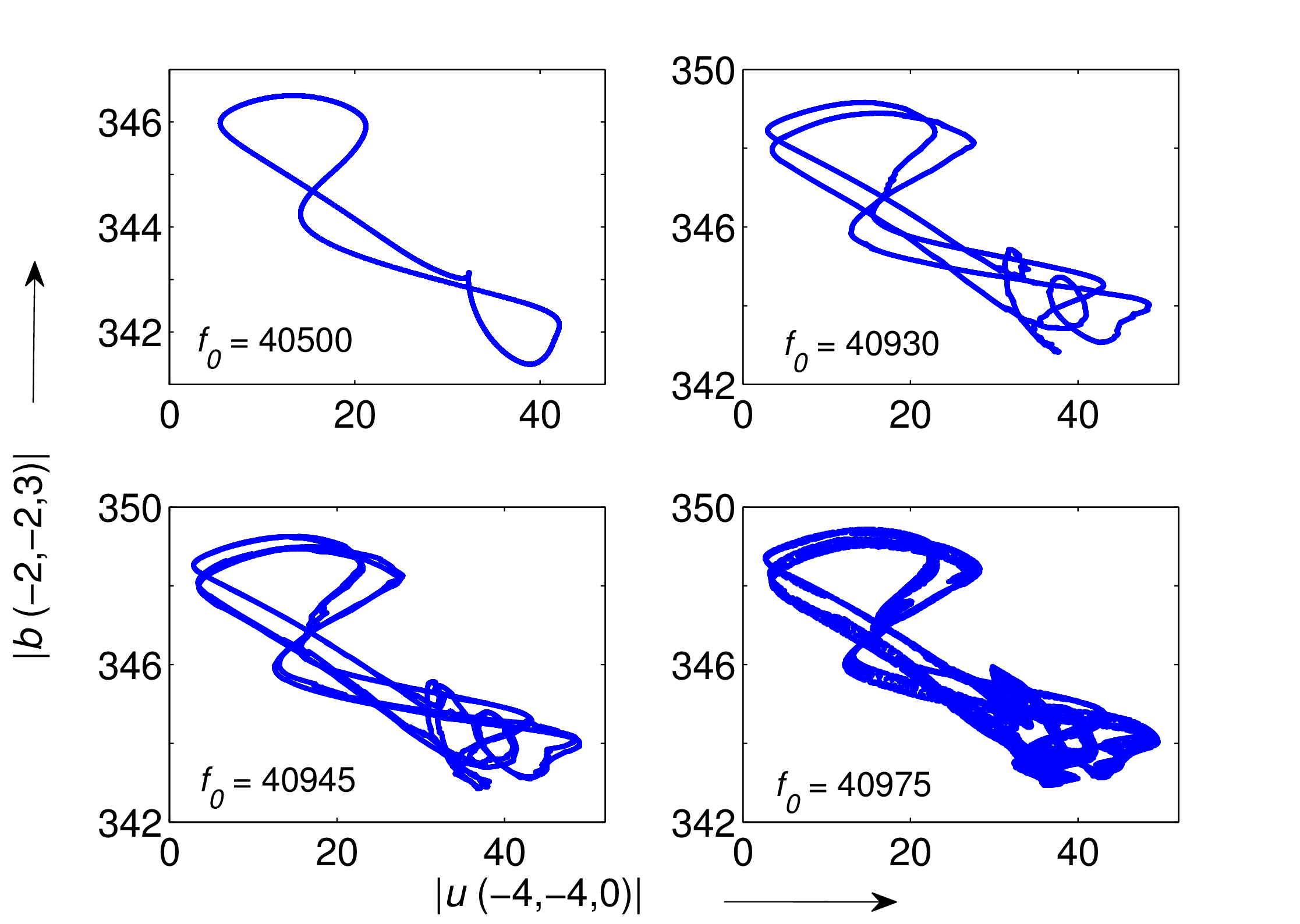}
\caption{For $\mathrm{Pm} = 1$: Poincar{\'e} sections taken at $|b_4|$ = mean($|b_4|$) for different forcing amplitude showing period doubling of the Poincar{\'e} map of the quasi-periodic state. For $f_0=40500$, the Poincar{\'e} map is of period one, for $f_0=40930$ period two, for $f_0=40945$ period four, and for $f_0=40975$ it is of higher periods.}
\label{fig:poincare}
\end{figure}

For forcing amplitude just above the critical forcing, we observe fixed point (or constant) dynamo state. If we further increase the forcing amplitude, we observe periodic oscillatory solution. We get quasi-periodic solution at higher forcing amplitudes.  In Fig.~\ref{fig:poincare}, we show the Poincar{\'e} sections taken at $|b_4|$ = mean($|b_4|$), which indicate a period-doubling of the Poincar{\'e} map of the quasi-periodic state as we increase the forcing amplitude. As we further increase the forcing, the quasi-periodic solution appears to turn into a chaotic state. In a low-dimensional model Donner {\it et al.}~\cite{Donner:PD2006} have also observed stationary, periodic, quasi-periodic, and chaotic magnetic fields with increasing Reynolds number. We have observed analytically in Sec.~\ref{sec:dyn_transition} that for the fixed point dynamo state, both $u_1$ and $u_2$ are real, $u_3$ is purely imaginary, and both the magnetic modes $b_4$ and $b_5$ are complex numbers. Our numerical simulations indicate that the same is true for the periodic dynamo states as well. In the case of quasi-periodic and chaotic solutions, all the five modes are complex numbers. 

\begin{figure}[htbp]
\centering
\includegraphics[scale=0.48]{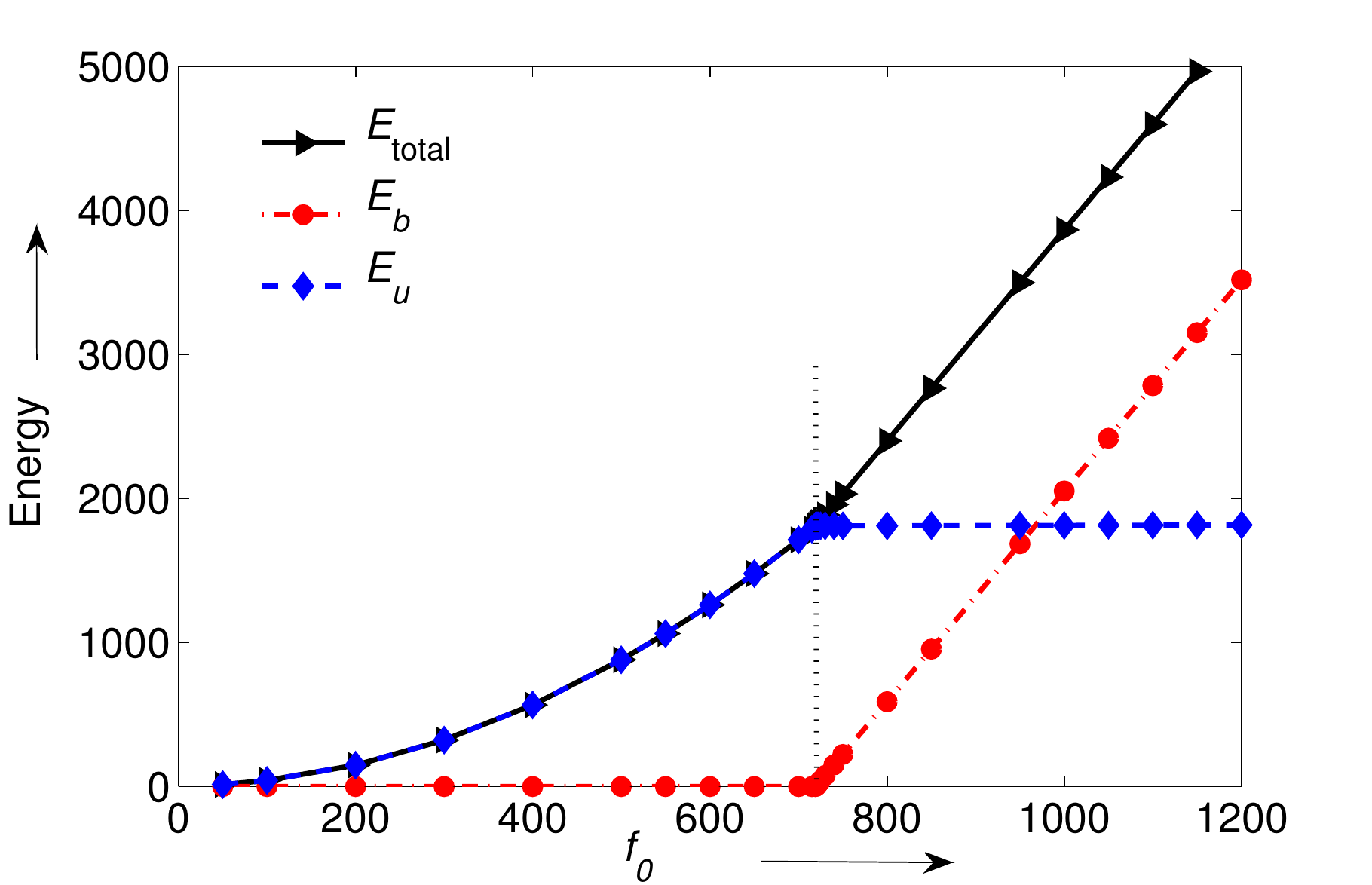}
\caption{For $\mathrm{Pm} = 1$: Plot of kinetic energy ($E_u$), magnetic energy ($E_b$), and total energy ($E_{total} = E_u + E_b$) below and above the onset of the dynamo ($f_c \approx 720$). After dynamo transition, the kinetic energy saturates to a constant value whereas the magnetic energy increases linearly.}
\label{fig:energy}
\end{figure}

In Fig.~\ref{fig:energy}, we show the variation of kinetic energy [$E_u = (|u_1|^2 + |u_2|^2 +|u_3|^2)/2$], magnetic energy [$E_b= (|b_4|^2 + |b_5|^2 )/2$], and total energy ($E_{total} = E_u + E_b$) with applied forcing, before and after the dynamo transition. The kinetic energy, which is the total energy before the dynamo transition, increases slowly in the regime below the critical forcing. As the dynamo excites, the kinetic energy saturates to a constant value and both the magnetic and total energies increase linearly. It indicates that after dynamo transition the magnitude of the velocity modes remain almost constant and the magnetic modes keep increasing. Note that in our model, $\mathrm{Pm} = 1$ falls in the category of low $\mathrm{Pm}$, and in this limit $u_2$ is the most dominant velocity mode. We observe that after the dynamo transition, $u_2$ saturates and in turn the kinetic energy saturates.   

We also estimate the efficiency of dynamo for $\mathrm{Pm}=1$ by calculating the ratio of magnetic energy and kinetic energy. This ratio increases continuously with the forcing amplitude and its value is greater than one (i.e., $E_b > E_u$) for higher values of forcing ($f_0 \sim 1000$). It implies that the magnetic modes dominate over kinetic modes above a certain forcing amplitude.

\begin{figure}[htbp]
\centering
\includegraphics[scale=0.11]{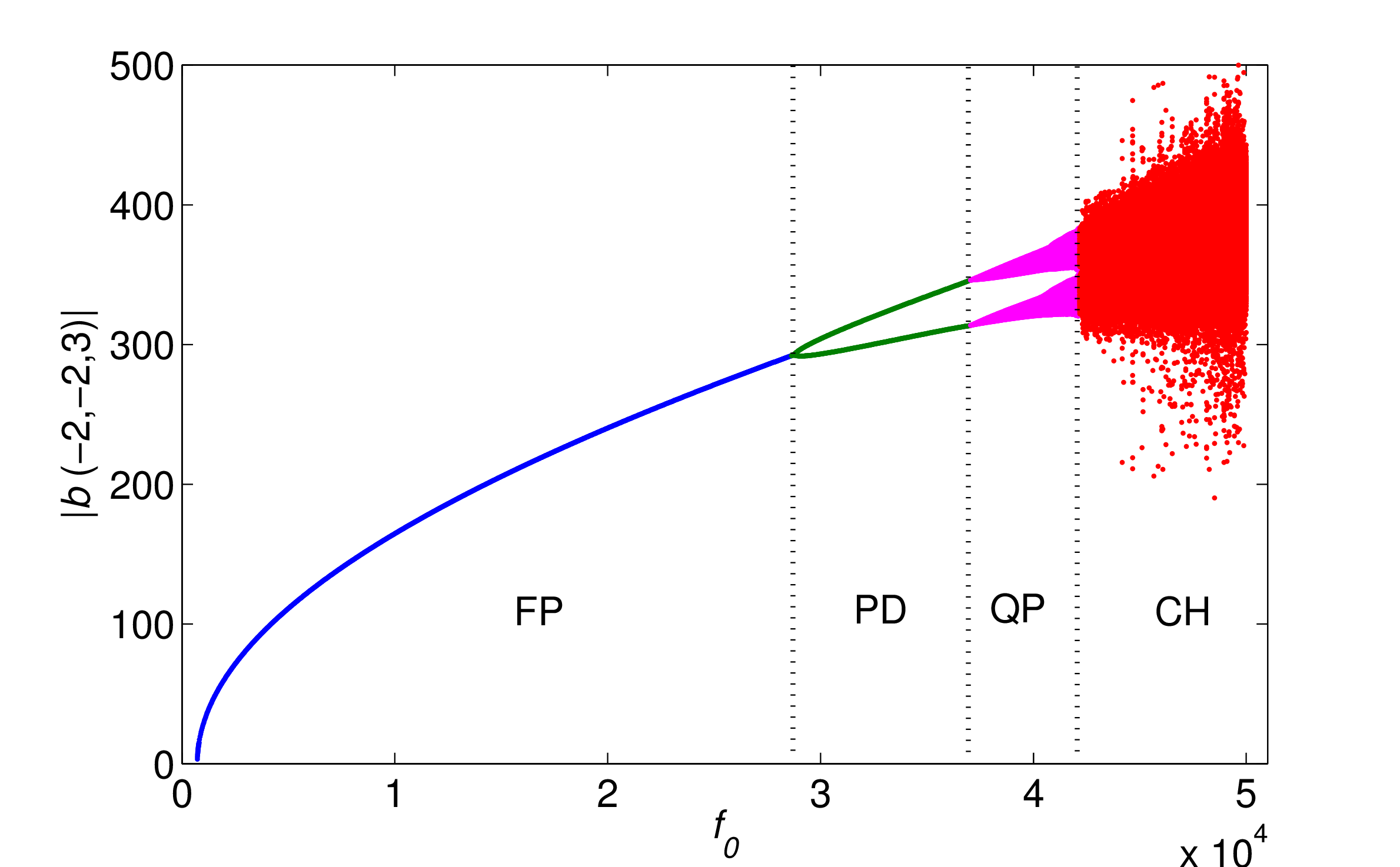}
\caption{For $\mathrm{Pm}=1$: Bifurcation diagram depicting the extremas of $|b(-2,-2,3)|$ with forcing amplitude. Stationary state or fixed point dynamo (FP) solution starts from $f_0 \approx 720$, periodic (PD) magnetic field is from $f_0 \approx 28700$, later quasi-periodic (QP) magnetic field appears from $f_0 \approx 37100$, and in the final stages the chaos (CH) appears from $f_0 \approx 42200$ and up to the later part of the plot.}
\label{fig:bif_diag}
\end{figure}

To get an idea of windows of different types of dynamo states, we show the overall bifurcation diagram depicting the extremas of $|b_5|$ as a function of the forcing amplitude in Fig.~\ref{fig:bif_diag}. We get stationary dynamo state just above the onset of dynamo (at $f_0 \approx 720$), and by increasing the forcing amplitude further, the dynamical system of equations produce periodic (at $f_0 \approx 28700$), quasi-periodic (at $f_0 \approx 37100$), and chaotic (at $f_0 \approx 42200$) dynamo solutions. In our low-dimensional model, chaos is achieved via a quasi-periodic route, which have been reported in DNS results.~\cite{Yadav:EPL2010}  We also remark here that we observe chaos for all the magnetic Prandtl numbers ($\mathrm{Pm} > 1$ and $\mathrm{Pm} < 1$) and it is through quasi-periodic route. We expect that the nature of the bifurcation diagrams for $\mathrm{Pm} = 100$, $0.01$, etc., would appear qualitatively similar to that of $\mathrm{Pm} =1$ and hence we do not show them here. Also, it is time consuming to plot the bifurcation diagrams for very small and very large $\mathrm{Pm}$.   Furthermore, for very small $\mathrm{Pm}$, the onset of dynamo is known to be  subcritical, whereas our model has a supercritical dynamo transition for all $\mathrm{Pm}$. Hence,  a refined model is required to capture the detailed bifurcation behavior of small $\mathrm{Pm}$ which has been left for future work.


\section{Discussion and Conclusions}
\label{sec:conclusion}

In this paper, we have presented a low-dimensional model by choosing five large-scale modes; three of which are velocity, and two are magnetic modes. We force two velocity modes to observe dynamo for very high and very low magnetic Prandtl numbers.  Analytical calculations show that the critical magnetic Reynolds number for dynamo ($\mathrm{Rm_c}$) saturates to a constant value in the two limiting cases of very high and very low magnetic Prandtl numbers. This result is important because performing DNS for such a broad range of $\mathrm{Pm}$ is unrealistic and low-dimensional model can be used to fill that gap. 

We find that in the case of very low $\mathrm{Pm}$, only one velocity mode is dominant, but for high $\mathrm{Pm}$ two velocity modes are significant and are almost equal in magnitude, whereas the third velocity mode is very small. These outcomes are possibly due to different kinds of triadic interactions in the two limiting cases of $\mathrm{Pm}$. We also observe that the critical forcing for dynamo decreases with increasing $\mathrm{Pm}$. 

The dynamo transition occurs through a supercritical pitchfork bifurcation of the fluid state. After dynamo transition for $\mathrm{Pm} =1$, the magnetic energy as well as the total energy increase linearly while kinetic energy remains almost constant. As we further increase the forcing amplitude, the magnetic energy dominates over the kinetic energy. For $\mathrm{Pm} =1$, as the forcing amplitude increased far above the critical forcing, we observe periodic, quasi-periodic, and chaotic dynamo states.

In summary, although this low-dimensional model reflects very few properties related to the dynamo, but, keeping in mind that performing DNS for such a vast range of $\mathrm{Pm}$ is impractical, some of the results observed through this model may be important and beneficial for further study of the dynamo. In this paper, we have mainly focused on the calculation of $\mathrm{Rm}_c$ for different $\mathrm{Pm}$ by using one set of possible (the trivial) solutions to the five-mode model, and have not considered some other possible solutions. A detailed analysis of this model with other solutions would be presented in a future work.


\begin{acknowledgments}
We thank T. Lessinnes for the suggestion to use this helical model; M. K. Verma, D. Carati, and R. Yadav for useful help, suggestions and comments. We are grateful to the anonymous referee for comments that helped us improve the manuscript. Rohit Kumar was partially supported by a research grant SERB/F/3279/2013-14 from Science and Engineering Research Board, India to M. K. Verma for which he is really thankful.
\end{acknowledgments}


\bibliography{turbulence}

\begin{thebibliography}{32}%
\makeatletter
\providecommand \@ifxundefined [1]{%
 \@ifx{#1\undefined}
}%
\providecommand \@ifnum [1]{%
 \ifnum #1\expandafter \@firstoftwo
 \else \expandafter \@secondoftwo
 \fi
}%
\providecommand \@ifx [1]{%
 \ifx #1\expandafter \@firstoftwo
 \else \expandafter \@secondoftwo
 \fi
}%
\providecommand \natexlab [1]{#1}%
\providecommand \enquote  [1]{``#1''}%
\providecommand \bibnamefont  [1]{#1}%
\providecommand \bibfnamefont [1]{#1}%
\providecommand \citenamefont [1]{#1}%
\providecommand \href@noop [0]{\@secondoftwo}%
\providecommand \href [0]{\begingroup \@sanitize@url \@href}%
\providecommand \@href[1]{\@@startlink{#1}\@@href}%
\providecommand \@@href[1]{\endgroup#1\@@endlink}%
\providecommand \@sanitize@url [0]{\catcode `\\12\catcode `\$12\catcode
  `\&12\catcode `\#12\catcode `\^12\catcode `\_12\catcode `\%12\relax}%
\providecommand \@@startlink[1]{}%
\providecommand \@@endlink[0]{}%
\providecommand \url  [0]{\begingroup\@sanitize@url \@url }%
\providecommand \@url [1]{\endgroup\@href {#1}{\urlprefix }}%
\providecommand \urlprefix  [0]{URL }%
\providecommand \Eprint [0]{\href }%
\providecommand \doibase [0]{http://dx.doi.org/}%
\providecommand \selectlanguage [0]{\@gobble}%
\providecommand \bibinfo  [0]{\@secondoftwo}%
\providecommand \bibfield  [0]{\@secondoftwo}%
\providecommand \translation [1]{[#1]}%
\providecommand \BibitemOpen [0]{}%
\providecommand \bibitemStop [0]{}%
\providecommand \bibitemNoStop [0]{.\EOS\space}%
\providecommand \EOS [0]{\spacefactor3000\relax}%
\providecommand \BibitemShut  [1]{\csname bibitem#1\endcsname}%
\let\auto@bib@innerbib\@empty
\bibitem [{\citenamefont {{Moffatt}}(1978)}]{Moffat:book}%
  \BibitemOpen
  \bibfield  {author} {\bibinfo {author} {\bibfnamefont {H.~K.}\ \bibnamefont
  {{Moffatt}}},\ }\href@noop {} {\emph {\bibinfo {title} {Magnetic Field
  Generation in Electrically Conducting Fluids}}}\ (\bibinfo  {publisher}
  {Cambridge university press},\ \bibinfo {address} {Cambridge},\ \bibinfo
  {year} {1978})\BibitemShut {NoStop}%
\bibitem [{\citenamefont {{Monchaux}}\ \emph {et~al.}(2007)\citenamefont
  {{Monchaux}}, \citenamefont {{Berhanu}}, \citenamefont {{Bourgoin}},
  \citenamefont {{Moulin}}, \citenamefont {{Odier}}, \citenamefont {{Pinton}},
  \citenamefont {{Volk}}, \citenamefont {{Fauve}}, \citenamefont {{Mordant}},
  \citenamefont {{P{\'e}tr{\'e}lis}}, \citenamefont {{Chiffaudel}},
  \citenamefont {{Daviaud}}, \citenamefont {{Dubrulle}}, \citenamefont
  {{Gasquet}}, \citenamefont {{Mari{\'e}}},\ and\ \citenamefont
  {{Ravelet}}}]{Monchaux:PRL2007}%
  \BibitemOpen
  \bibfield  {author} {\bibinfo {author} {\bibfnamefont {R.}~\bibnamefont
  {{Monchaux}}}, \bibinfo {author} {\bibfnamefont {M.}~\bibnamefont
  {{Berhanu}}}, \bibinfo {author} {\bibfnamefont {M.}~\bibnamefont
  {{Bourgoin}}}, \bibinfo {author} {\bibfnamefont {M.}~\bibnamefont
  {{Moulin}}}, \bibinfo {author} {\bibfnamefont {P.}~\bibnamefont {{Odier}}},
  \bibinfo {author} {\bibfnamefont {J.~F.}\ \bibnamefont {{Pinton}}}, \bibinfo
  {author} {\bibfnamefont {R.}~\bibnamefont {{Volk}}}, \bibinfo {author}
  {\bibfnamefont {S.}~\bibnamefont {{Fauve}}}, \bibinfo {author} {\bibfnamefont
  {N.}~\bibnamefont {{Mordant}}}, \bibinfo {author} {\bibfnamefont
  {F.}~\bibnamefont {{P{\'e}tr{\'e}lis}}}, \bibinfo {author} {\bibfnamefont
  {A.}~\bibnamefont {{Chiffaudel}}}, \bibinfo {author} {\bibfnamefont
  {F.}~\bibnamefont {{Daviaud}}}, \bibinfo {author} {\bibfnamefont
  {B.}~\bibnamefont {{Dubrulle}}}, \bibinfo {author} {\bibfnamefont
  {C.}~\bibnamefont {{Gasquet}}}, \bibinfo {author} {\bibfnamefont
  {L.}~\bibnamefont {{Mari{\'e}}}}, \ and\ \bibinfo {author} {\bibfnamefont
  {F.}~\bibnamefont {{Ravelet}}},\ }\href {\doibase
  10.1103/PhysRevLett.98.044502} {\bibfield  {journal} {\bibinfo  {journal}
  {Phys. Rev. Lett.}\ }\textbf {\bibinfo {volume} {98}},\ \bibinfo {pages}
  {044502} (\bibinfo {year} {2007})}\BibitemShut {NoStop}%
\bibitem [{\citenamefont {{Ponty}}\ and\ \citenamefont
  {{Politano}}(2004)}]{Ponty:PRL2004}%
  \BibitemOpen
  \bibfield  {author} {\bibinfo {author} {\bibfnamefont {Y.}~\bibnamefont
  {{Ponty}}}\ and\ \bibinfo {author} {\bibfnamefont {H.}~\bibnamefont
  {{Politano}}},\ }\href {\doibase 10.1103/PhysRevLett.92.144503} {\bibfield
  {journal} {\bibinfo  {journal} {Phys. Rev. Lett.}\ }\textbf {\bibinfo
  {volume} {92}},\ \bibinfo {pages} {144503} (\bibinfo {year}
  {2004})}\BibitemShut {NoStop}%
\bibitem [{\citenamefont {{Kazantsev}}(1968)}]{Kazantsev:JETP1968}%
  \BibitemOpen
  \bibfield  {author} {\bibinfo {author} {\bibfnamefont {A.~P.}\ \bibnamefont
  {{Kazantsev}}},\ }\href@noop {} {\bibfield  {journal} {\bibinfo  {journal}
  {Sov. Phys. JETP}\ }\textbf {\bibinfo {volume} {26}},\ \bibinfo {pages}
  {1031} (\bibinfo {year} {1968})}\BibitemShut {NoStop}%
\bibitem [{\citenamefont {{Brandenburg}}\ and\ \citenamefont
  {{Subramanian}}(2005)}]{Brandenburg:PR2005}%
  \BibitemOpen
  \bibfield  {author} {\bibinfo {author} {\bibfnamefont {A.}~\bibnamefont
  {{Brandenburg}}}\ and\ \bibinfo {author} {\bibfnamefont {K.}~\bibnamefont
  {{Subramanian}}},\ }\href {\doibase 10.1016/j.physrep.2005.06.005} {\bibfield
   {journal} {\bibinfo  {journal} {Phys. Rep.}\ }\textbf {\bibinfo {volume}
  {417}},\ \bibinfo {pages} {1} (\bibinfo {year} {2005})}\BibitemShut {NoStop}%
\bibitem [{\citenamefont {{Plunian}}, \citenamefont {{Stepanov}},\ and\
  \citenamefont {{Frick}}(2013)}]{Plunian:PR2013}%
  \BibitemOpen
  \bibfield  {author} {\bibinfo {author} {\bibfnamefont {F.}~\bibnamefont
  {{Plunian}}}, \bibinfo {author} {\bibfnamefont {R.}~\bibnamefont
  {{Stepanov}}}, \ and\ \bibinfo {author} {\bibfnamefont {P.}~\bibnamefont
  {{Frick}}},\ }\href {\doibase 10.1016/j.physrep.2012.09.001} {\bibfield
  {journal} {\bibinfo  {journal} {Phys. Rep.}\ }\textbf {\bibinfo {volume}
  {523}},\ \bibinfo {pages} {1} (\bibinfo {year} {2013})}\BibitemShut {NoStop}%
\bibitem [{\citenamefont {{Glatzmaier}}\ and\ \citenamefont
  {{Roberts}}(1995)}]{Glatzmaier:NATURE1995}%
  \BibitemOpen
  \bibfield  {author} {\bibinfo {author} {\bibfnamefont {G.}~\bibnamefont
  {{Glatzmaier}}}\ and\ \bibinfo {author} {\bibfnamefont {P.}~\bibnamefont
  {{Roberts}}},\ }\href@noop {} {\bibfield  {journal} {\bibinfo  {journal}
  {Nature}\ }\textbf {\bibinfo {volume} {377}},\ \bibinfo {pages} {203}
  (\bibinfo {year} {1995})}\BibitemShut {NoStop}%
\bibitem [{\citenamefont {{Ponty}}\ \emph {et~al.}(2007)\citenamefont
  {{Ponty}}, \citenamefont {{Laval}}, \citenamefont {{Dubrulle}}, \citenamefont
  {{Daviaud}},\ and\ \citenamefont {{Pinton}}}]{Ponty:PRL2007}%
  \BibitemOpen
  \bibfield  {author} {\bibinfo {author} {\bibfnamefont {Y.}~\bibnamefont
  {{Ponty}}}, \bibinfo {author} {\bibfnamefont {J.}~\bibnamefont {{Laval}}},
  \bibinfo {author} {\bibfnamefont {B.}~\bibnamefont {{Dubrulle}}}, \bibinfo
  {author} {\bibfnamefont {F.}~\bibnamefont {{Daviaud}}}, \ and\ \bibinfo
  {author} {\bibfnamefont {J.}~\bibnamefont {{Pinton}}},\ }\href {\doibase
  10.1103/PhysRevLett.99.224501} {\bibfield  {journal} {\bibinfo  {journal}
  {Phys. Rev. Lett.}\ }\textbf {\bibinfo {volume} {99}},\ \bibinfo {pages}
  {224501} (\bibinfo {year} {2007})}\BibitemShut {NoStop}%
\bibitem [{\citenamefont {{Mininni}}, \citenamefont {{Montgomery}},\ and\
  \citenamefont {{Pouquet}}(2005)}]{Mininni:PRE2005}%
  \BibitemOpen
  \bibfield  {author} {\bibinfo {author} {\bibfnamefont {P.}~\bibnamefont
  {{Mininni}}}, \bibinfo {author} {\bibfnamefont {D.}~\bibnamefont
  {{Montgomery}}}, \ and\ \bibinfo {author} {\bibfnamefont {A.}~\bibnamefont
  {{Pouquet}}},\ }\href {\doibase 10.1103/PhysRevE.71.046304} {\bibfield
  {journal} {\bibinfo  {journal} {Phys. Rev. E}\ }\textbf {\bibinfo {volume}
  {71}},\ \bibinfo {pages} {046304} (\bibinfo {year} {2005})}\BibitemShut
  {NoStop}%
\bibitem [{\citenamefont {{Ravelet}}, \citenamefont {{Chiffaudel}},\ and\
  \citenamefont {{Daviaud}}(2005)}]{Ravelet:PF2005}%
  \BibitemOpen
  \bibfield  {author} {\bibinfo {author} {\bibfnamefont {F.}~\bibnamefont
  {{Ravelet}}}, \bibinfo {author} {\bibfnamefont {A.}~\bibnamefont
  {{Chiffaudel}}}, \ and\ \bibinfo {author} {\bibfnamefont {F.}~\bibnamefont
  {{Daviaud}}},\ }\href {\doibase 10.1063/1.2130745} {\bibfield  {journal}
  {\bibinfo  {journal} {Phys. Fluids}\ }\textbf {\bibinfo {volume} {17}},\
  \bibinfo {pages} {117104} (\bibinfo {year} {2005})}\BibitemShut {NoStop}%
\bibitem [{\citenamefont {{Yousef}}\ \emph {et~al.}(2008)\citenamefont
  {{Yousef}}, \citenamefont {{Heinemann}}, \citenamefont {{Schekochihin}},
  \citenamefont {{Kleeorin}}, \citenamefont {{Rogachevskii}}, \citenamefont
  {{Iskakov}}, \citenamefont {{Cowley}},\ and\ \citenamefont
  {{McWilliams}}}]{Yousef:PRL2008}%
  \BibitemOpen
  \bibfield  {author} {\bibinfo {author} {\bibfnamefont {T.~A.}\ \bibnamefont
  {{Yousef}}}, \bibinfo {author} {\bibfnamefont {T.}~\bibnamefont
  {{Heinemann}}}, \bibinfo {author} {\bibfnamefont {A.~A.}\ \bibnamefont
  {{Schekochihin}}}, \bibinfo {author} {\bibfnamefont {N.}~\bibnamefont
  {{Kleeorin}}}, \bibinfo {author} {\bibfnamefont {I.}~\bibnamefont
  {{Rogachevskii}}}, \bibinfo {author} {\bibfnamefont {A.~B.}\ \bibnamefont
  {{Iskakov}}}, \bibinfo {author} {\bibfnamefont {S.~C.}\ \bibnamefont
  {{Cowley}}}, \ and\ \bibinfo {author} {\bibfnamefont {J.~C.}\ \bibnamefont
  {{McWilliams}}},\ }\href {\doibase 10.1103/PhysRevLett.100.184501} {\bibfield
   {journal} {\bibinfo  {journal} {Phys. Rev. Lett.}\ }\textbf {\bibinfo
  {volume} {100}},\ \bibinfo {pages} {184501} (\bibinfo {year}
  {2008})}\BibitemShut {NoStop}%
\bibitem [{\citenamefont {{Yadav}}\ \emph {et~al.}(2010)\citenamefont
  {{Yadav}}, \citenamefont {{Chandra}}, \citenamefont {{Verma}}, \citenamefont
  {{Paul}},\ and\ \citenamefont {{Wahi}}}]{Yadav:EPL2010}%
  \BibitemOpen
  \bibfield  {author} {\bibinfo {author} {\bibfnamefont {R.}~\bibnamefont
  {{Yadav}}}, \bibinfo {author} {\bibfnamefont {M.}~\bibnamefont {{Chandra}}},
  \bibinfo {author} {\bibfnamefont {M.~K.}\ \bibnamefont {{Verma}}}, \bibinfo
  {author} {\bibfnamefont {S.}~\bibnamefont {{Paul}}}, \ and\ \bibinfo {author}
  {\bibfnamefont {P.}~\bibnamefont {{Wahi}}},\ }\href {\doibase
  10.1209/0295-5075/91/69001} {\bibfield  {journal} {\bibinfo  {journal}
  {Europhys. Lett.}\ }\textbf {\bibinfo {volume} {91}},\ \bibinfo {pages}
  {69001} (\bibinfo {year} {2010})}\BibitemShut {NoStop}%
\bibitem [{\citenamefont {{Kumar}}, \citenamefont {{Verma}},\ and\
  \citenamefont {{Samtaney}}(2013)}]{Kumar:EPL2013}%
  \BibitemOpen
  \bibfield  {author} {\bibinfo {author} {\bibfnamefont {R.}~\bibnamefont
  {{Kumar}}}, \bibinfo {author} {\bibfnamefont {M.~K.}\ \bibnamefont
  {{Verma}}}, \ and\ \bibinfo {author} {\bibfnamefont {R.}~\bibnamefont
  {{Samtaney}}},\ }\href {\doibase 10.1209/0295-5075/104/54001} {\bibfield
  {journal} {\bibinfo  {journal} {Europhys. Lett.}\ }\textbf {\bibinfo {volume}
  {104}},\ \bibinfo {pages} {54001} (\bibinfo {year} {2013})}\BibitemShut
  {NoStop}%
\bibitem [{\citenamefont {{Kumar}}, \citenamefont {{Verma}},\ and\
  \citenamefont {{Samtaney}}(2015)}]{Kumar:JT2015}%
  \BibitemOpen
  \bibfield  {author} {\bibinfo {author} {\bibfnamefont {R.}~\bibnamefont
  {{Kumar}}}, \bibinfo {author} {\bibfnamefont {M.~K.}\ \bibnamefont
  {{Verma}}}, \ and\ \bibinfo {author} {\bibfnamefont {R.}~\bibnamefont
  {{Samtaney}}},\ }\href {\doibase 10.1080/14685248.2015.1057340} {\bibfield
  {journal} {\bibinfo  {journal} {J. Turbul.}\ }\textbf {\bibinfo {volume}
  {16}},\ \bibinfo {pages} {1114} (\bibinfo {year} {2015})}\BibitemShut
  {NoStop}%
\bibitem [{\citenamefont {{Ponty}}\ \emph {et~al.}(2005)\citenamefont
  {{Ponty}}, \citenamefont {{Mininni}}, \citenamefont {{Montgomery}},
  \citenamefont {{Pinton}}, \citenamefont {{Politano}},\ and\ \citenamefont
  {{Pouquet}}}]{Ponty:PRL2005}%
  \BibitemOpen
  \bibfield  {author} {\bibinfo {author} {\bibfnamefont {Y.}~\bibnamefont
  {{Ponty}}}, \bibinfo {author} {\bibfnamefont {P.~D.}\ \bibnamefont
  {{Mininni}}}, \bibinfo {author} {\bibfnamefont {D.~C.}\ \bibnamefont
  {{Montgomery}}}, \bibinfo {author} {\bibfnamefont {J.~F.}\ \bibnamefont
  {{Pinton}}}, \bibinfo {author} {\bibfnamefont {H.}~\bibnamefont
  {{Politano}}}, \ and\ \bibinfo {author} {\bibfnamefont {A.}~\bibnamefont
  {{Pouquet}}},\ }\href {\doibase 10.1103/PhysRevLett.94.164502} {\bibfield
  {journal} {\bibinfo  {journal} {Phys. Rev. Lett.}\ }\textbf {\bibinfo
  {volume} {94}},\ \bibinfo {pages} {164502} (\bibinfo {year}
  {2005})}\BibitemShut {NoStop}%
\bibitem [{\citenamefont {{Haugen}}, \citenamefont {{Brandenburg}},\ and\
  \citenamefont {{Dobler}}(2004)}]{Haugen:PRE2004}%
  \BibitemOpen
  \bibfield  {author} {\bibinfo {author} {\bibfnamefont {N.~E.}\ \bibnamefont
  {{Haugen}}}, \bibinfo {author} {\bibfnamefont {A.}~\bibnamefont
  {{Brandenburg}}}, \ and\ \bibinfo {author} {\bibfnamefont {W.}~\bibnamefont
  {{Dobler}}},\ }\href {\doibase 10.1103/PhysRevE.70.016308} {\bibfield
  {journal} {\bibinfo  {journal} {Phys. Rev. E}\ }\textbf {\bibinfo {volume}
  {70}},\ \bibinfo {pages} {016308} (\bibinfo {year} {2004})}\BibitemShut
  {NoStop}%
\bibitem [{\citenamefont {{Schekochihin}}\ \emph
  {et~al.}(2004{\natexlab{a}})\citenamefont {{Schekochihin}}, \citenamefont
  {{Cowley}}, \citenamefont {{Taylor}}, \citenamefont {{Maron}},\ and\
  \citenamefont {{McWilliams}}}]{Schekochihin:APJ2004}%
  \BibitemOpen
  \bibfield  {author} {\bibinfo {author} {\bibfnamefont {A.~A.}\ \bibnamefont
  {{Schekochihin}}}, \bibinfo {author} {\bibfnamefont {S.~C.}\ \bibnamefont
  {{Cowley}}}, \bibinfo {author} {\bibfnamefont {S.~F.}\ \bibnamefont
  {{Taylor}}}, \bibinfo {author} {\bibfnamefont {J.~L.}\ \bibnamefont
  {{Maron}}}, \ and\ \bibinfo {author} {\bibfnamefont {J.~C.}\ \bibnamefont
  {{McWilliams}}},\ }\href {\doibase 10.1086/422547} {\bibfield  {journal}
  {\bibinfo  {journal} {Astrophys. J.}\ }\textbf {\bibinfo {volume} {612}},\
  \bibinfo {pages} {276} (\bibinfo {year} {2004}{\natexlab{a}})}\BibitemShut
  {NoStop}%
\bibitem [{\citenamefont {{Schekochihin}}\ \emph
  {et~al.}(2004{\natexlab{b}})\citenamefont {{Schekochihin}}, \citenamefont
  {{Cowley}}, \citenamefont {{Maron}},\ and\ \citenamefont
  {{Mcwilliamss}}}]{Schekochihin:PRL2004b}%
  \BibitemOpen
  \bibfield  {author} {\bibinfo {author} {\bibfnamefont {A.~A.}\ \bibnamefont
  {{Schekochihin}}}, \bibinfo {author} {\bibfnamefont {S.~C.}\ \bibnamefont
  {{Cowley}}}, \bibinfo {author} {\bibfnamefont {J.~L.}\ \bibnamefont
  {{Maron}}}, \ and\ \bibinfo {author} {\bibfnamefont {J.~C.}\ \bibnamefont
  {{Mcwilliamss}}},\ }\href {\doibase 10.1103/PhysRevLett.92.054502} {\bibfield
   {journal} {\bibinfo  {journal} {Phys. Rev. Lett.}\ }\textbf {\bibinfo
  {volume} {92}},\ \bibinfo {pages} {054502} (\bibinfo {year}
  {2004}{\natexlab{b}})}\BibitemShut {NoStop}%
\bibitem [{\citenamefont {{Rikitake}}(1958)}]{Rikitake:MPTCPS1958}%
  \BibitemOpen
  \bibfield  {author} {\bibinfo {author} {\bibfnamefont {T.}~\bibnamefont
  {{Rikitake}}},\ }\href {\doibase 10.1017/S0305004100033223} {\bibfield
  {journal} {\bibinfo  {journal} {Mathematical proceedings of the Cambridge
  philosophical society}\ }\textbf {\bibinfo {volume} {54}},\ \bibinfo {pages}
  {89} (\bibinfo {year} {1958})}\BibitemShut {NoStop}%
\bibitem [{\citenamefont {{Gissinger}}, \citenamefont {{Dormy}},\ and\
  \citenamefont {{Fauve}}(2010)}]{Gissinger:EPL2010}%
  \BibitemOpen
  \bibfield  {author} {\bibinfo {author} {\bibfnamefont {C.}~\bibnamefont
  {{Gissinger}}}, \bibinfo {author} {\bibfnamefont {E.}~\bibnamefont
  {{Dormy}}}, \ and\ \bibinfo {author} {\bibfnamefont {S.}~\bibnamefont
  {{Fauve}}},\ }\href {\doibase 10.1209/0295-5075/90/49001} {\bibfield
  {journal} {\bibinfo  {journal} {EPL}\ }\textbf {\bibinfo {volume} {90}},\
  \bibinfo {pages} {49001} (\bibinfo {year} {2010})}\BibitemShut {NoStop}%
\bibitem [{\citenamefont {{Gissinger}}(2012)}]{Gissinger:EPJB2012}%
  \BibitemOpen
  \bibfield  {author} {\bibinfo {author} {\bibfnamefont {C.}~\bibnamefont
  {{Gissinger}}},\ }\href {\doibase 10.1140/epjb/e2012-20799-5} {\bibfield
  {journal} {\bibinfo  {journal} {Eur. Phys. J. B}\ }\textbf {\bibinfo {volume}
  {85}},\ \bibinfo {pages} {137} (\bibinfo {year} {2012})}\BibitemShut
  {NoStop}%
\bibitem [{\citenamefont {{Donner}}\ \emph {et~al.}(2006)\citenamefont
  {{Donner}}, \citenamefont {{Seehafer}}, \citenamefont {{Sanju\'{a}n}},\ and\
  \citenamefont {{Feudel}}}]{Donner:PD2006}%
  \BibitemOpen
  \bibfield  {author} {\bibinfo {author} {\bibfnamefont {R.}~\bibnamefont
  {{Donner}}}, \bibinfo {author} {\bibfnamefont {N.}~\bibnamefont
  {{Seehafer}}}, \bibinfo {author} {\bibfnamefont {M.~A.}\ \bibnamefont
  {{Sanju\'{a}n}}}, \ and\ \bibinfo {author} {\bibfnamefont {F.}~\bibnamefont
  {{Feudel}}},\ }\href {\doibase 10.1016/j.physd.2006.08.022} {\bibfield
  {journal} {\bibinfo  {journal} {Physica D: Nonlinear Phenomena}\ }\textbf
  {\bibinfo {volume} {223}},\ \bibinfo {pages} {151} (\bibinfo {year}
  {2006})}\BibitemShut {NoStop}%
\bibitem [{\citenamefont {{Verma}}\ \emph {et~al.}(2008)\citenamefont
  {{Verma}}, \citenamefont {{Lessinnes}}, \citenamefont {{Carati}},
  \citenamefont {{Sarris}}, \citenamefont {{Kumar}},\ and\ \citenamefont
  {{Singh}}}]{Verma:PRE2008}%
  \BibitemOpen
  \bibfield  {author} {\bibinfo {author} {\bibfnamefont {M.~K.}\ \bibnamefont
  {{Verma}}}, \bibinfo {author} {\bibfnamefont {T.}~\bibnamefont
  {{Lessinnes}}}, \bibinfo {author} {\bibfnamefont {D.}~\bibnamefont
  {{Carati}}}, \bibinfo {author} {\bibfnamefont {I.}~\bibnamefont {{Sarris}}},
  \bibinfo {author} {\bibfnamefont {K.}~\bibnamefont {{Kumar}}}, \ and\
  \bibinfo {author} {\bibfnamefont {M.}~\bibnamefont {{Singh}}},\ }\href
  {\doibase 10.1103/PhysRevE.78.036409} {\bibfield  {journal} {\bibinfo
  {journal} {Phys. Rev. E}\ }\textbf {\bibinfo {volume} {78}},\ \bibinfo
  {pages} {036409} (\bibinfo {year} {2008})}\BibitemShut {NoStop}%
\bibitem [{\citenamefont {{Verma}}\ and\ \citenamefont
  {{Yadav}}(2013)}]{Verma:PP2013}%
  \BibitemOpen
  \bibfield  {author} {\bibinfo {author} {\bibfnamefont {M.~K.}\ \bibnamefont
  {{Verma}}}\ and\ \bibinfo {author} {\bibfnamefont {R.~K.}\ \bibnamefont
  {{Yadav}}},\ }\href {\doibase 10.1063/1.4813261} {\bibfield  {journal}
  {\bibinfo  {journal} {Phys. Plasmas}\ }\textbf {\bibinfo {volume} {20}},\
  \bibinfo {pages} {072307} (\bibinfo {year} {2013})}\BibitemShut {NoStop}%
\bibitem [{\citenamefont {{Verma}}(2004)}]{Verma:PR2004}%
  \BibitemOpen
  \bibfield  {author} {\bibinfo {author} {\bibfnamefont {M.~K.}\ \bibnamefont
  {{Verma}}},\ }\href {\doibase 10.1016/j.physrep.2004.07.007} {\bibfield
  {journal} {\bibinfo  {journal} {Phys. Rep.}\ }\textbf {\bibinfo {volume}
  {401}},\ \bibinfo {pages} {229} (\bibinfo {year} {2004})}\BibitemShut
  {NoStop}%
\bibitem [{\citenamefont {{Waleffe}}(1992)}]{waleffe:PF1992}%
  \BibitemOpen
  \bibfield  {author} {\bibinfo {author} {\bibfnamefont {F.}~\bibnamefont
  {{Waleffe}}},\ }\href {\doibase 10.1063/1.858309} {\bibfield  {journal}
  {\bibinfo  {journal} {Phys. Fluids A}\ }\textbf {\bibinfo {volume} {4}},\
  \bibinfo {pages} {350} (\bibinfo {year} {1992})}\BibitemShut {NoStop}%
\bibitem [{\citenamefont {Lessinnes}, \citenamefont {Plunian},\ and\
  \citenamefont {Carati}(2009)}]{Lessinnes:TCFD2009}%
  \BibitemOpen
  \bibfield  {author} {\bibinfo {author} {\bibfnamefont {T.}~\bibnamefont
  {Lessinnes}}, \bibinfo {author} {\bibfnamefont {F.}~\bibnamefont {Plunian}},
  \ and\ \bibinfo {author} {\bibfnamefont {D.}~\bibnamefont {Carati}},\ }\href
  {\doibase 10.1007/s00162-009-0165-y} {\bibfield  {journal} {\bibinfo
  {journal} {Theoretical and Computational Fluid Dynamics}\ }\textbf {\bibinfo
  {volume} {23}},\ \bibinfo {pages} {439} (\bibinfo {year} {2009})}\BibitemShut
  {NoStop}%
\bibitem [{\citenamefont {{Nigro}}\ and\ \citenamefont
  {{Veltri}}(2011)}]{Nigro:APJ2011}%
  \BibitemOpen
  \bibfield  {author} {\bibinfo {author} {\bibfnamefont {G.}~\bibnamefont
  {{Nigro}}}\ and\ \bibinfo {author} {\bibfnamefont {P.}~\bibnamefont
  {{Veltri}}},\ }\href {\doibase 10.1088/2041-8205/740/2/L37} {\bibfield
  {journal} {\bibinfo  {journal} {Astrophys. J.}\ }\textbf {\bibinfo {volume}
  {740}},\ \bibinfo {pages} {L37} (\bibinfo {year} {2011})}\BibitemShut
  {NoStop}%
\bibitem [{\citenamefont {{Yadav}}, \citenamefont {{Verma}},\ and\
  \citenamefont {{Wahi}}(2012)}]{Yadav:PRE2012}%
  \BibitemOpen
  \bibfield  {author} {\bibinfo {author} {\bibfnamefont {R.}~\bibnamefont
  {{Yadav}}}, \bibinfo {author} {\bibfnamefont {M.~K.}\ \bibnamefont
  {{Verma}}}, \ and\ \bibinfo {author} {\bibfnamefont {P.}~\bibnamefont
  {{Wahi}}},\ }\href {\doibase 10.1103/PhysRevE.85.036301} {\bibfield
  {journal} {\bibinfo  {journal} {Phys. Rev. E}\ }\textbf {\bibinfo {volume}
  {85}},\ \bibinfo {pages} {036301} (\bibinfo {year} {2012})}\BibitemShut
  {NoStop}%
\bibitem [{\citenamefont {{Krstulovic}}\ \emph {et~al.}(2011)\citenamefont
  {{Krstulovic}}, \citenamefont {{Thorner}}, \citenamefont {{Vest}},
  \citenamefont {{Fauve}},\ and\ \citenamefont
  {{Brachet}}}]{Krstulovic:PRE2011}%
  \BibitemOpen
  \bibfield  {author} {\bibinfo {author} {\bibfnamefont {G.}~\bibnamefont
  {{Krstulovic}}}, \bibinfo {author} {\bibfnamefont {G.}~\bibnamefont
  {{Thorner}}}, \bibinfo {author} {\bibfnamefont {J.}~\bibnamefont {{Vest}}},
  \bibinfo {author} {\bibfnamefont {S.}~\bibnamefont {{Fauve}}}, \ and\
  \bibinfo {author} {\bibfnamefont {M.}~\bibnamefont {{Brachet}}},\ }\href
  {\doibase 10.1103/PhysRevE.84.066318} {\bibfield  {journal} {\bibinfo
  {journal} {Phys. Rev. E}\ }\textbf {\bibinfo {volume} {84}},\ \bibinfo
  {pages} {066318} (\bibinfo {year} {2011})}\BibitemShut {NoStop}%
\bibitem [{\citenamefont {{Morin}}\ and\ \citenamefont
  {{Dormy}}(2009)}]{Morin:IJMPB2009}%
  \BibitemOpen
  \bibfield  {author} {\bibinfo {author} {\bibfnamefont {V.}~\bibnamefont
  {{Morin}}}\ and\ \bibinfo {author} {\bibfnamefont {E.}~\bibnamefont
  {{Dormy}}},\ }\href {\doibase 10.1142/S021797920906378X} {\bibfield
  {journal} {\bibinfo  {journal} {Int. J. Mod. Phys. B}\ }\textbf {\bibinfo
  {volume} {23}},\ \bibinfo {pages} {5467} (\bibinfo {year}
  {2009})}\BibitemShut {NoStop}%
\bibitem [{\citenamefont {{P\'etr\'elis}}\ and\ \citenamefont
  {{Fauve}}(2001)}]{Petrelis:EPJB2001}%
  \BibitemOpen
  \bibfield  {author} {\bibinfo {author} {\bibfnamefont {F.}~\bibnamefont
  {{P\'etr\'elis}}}\ and\ \bibinfo {author} {\bibfnamefont {S.}~\bibnamefont
  {{Fauve}}},\ }\href {\doibase 10.1007/s100510170103} {\bibfield  {journal}
  {\bibinfo  {journal} {Eur. Phys. J. B}\ }\textbf {\bibinfo {volume} {22}},\
  \bibinfo {pages} {273} (\bibinfo {year} {2001})}\BibitemShut {NoStop}%
\end{thebibliography}


%


\end{document}